\documentclass[12pt]{article}
\usepackage{epsfig,amssymb}

\newcommand\beq{\begin{equation}}
\newcommand\eeq{\end{equation}}
\newcommand\bea{\begin{eqnarray}}
\newcommand\eea{\end{eqnarray}}
\newcommand{\nonum}{\nonumber}
\newcommand\sqpi{\sqrt{\pi}}
\newcommand\px{\partial_x}
\newcommand\pt{\partial_\tau}
\newcommand\bo{{\bar \omega}}
\newcommand\on{{\bar \omega}_n}
\newcommand\tphi{\tilde\phi}
\newcommand\tchi{\tilde\chi}
\newcommand\ua{\uparrow}
\newcommand\da{\downarrow}

\newcommand\ttheta{\tilde\theta}
\newcommand\tk{\tilde k}

\begin{document}

\centerline{\Large Conductance through contact barriers of a finite length 
quantum wire}

\vskip .5 true cm
\centerline{\bf Siddhartha Lal$^1$, Sumathi Rao$^2$ and Diptiman Sen$^1$} 
\vskip .5 true cm

\centerline{\it $^1$ Centre for Theoretical Studies} 
\centerline{\it Indian Institute of Science, Bangalore 560012, India} 
\vskip .5 true cm

\centerline{\it $^2$ Harish-Chandra Research Institute}
\centerline{\it Chhatnag Road, Jhusi, Allahabad 211019, India}
\vskip .5 true cm

\begin{abstract}

We use the technique of bosonization to understand a variety of recent 
experimental results on the conductivity of a quantum wire. The quantum wire 
is taken to be a finite-length Luttinger liquid connected on two sides to 
semi-infinite Fermi liquids through contacts. The contacts are modeled as 
(short) Luttinger liquids bounded by localized one-body potentials. We use 
effective actions and the renormalization group to study the effects of
electronic interactions within the wire, the length of the wire, 
finite temperature and a magnetic field on the conductivity. We explain the 
deviations of the conductivity away from $2Ne^2 /h$ in wires which are not
too short as arising from renormalization effects caused by the repulsive 
interactions. We also explain the universal conductance corrections observed
in different channels at higher temperatures. We study the effects of an 
external magnetic field on electronic transport through this system and 
explain why odd and even spin split bands show different renormalizations 
from the universal conductance values. We discuss the case of resonant 
transmission and of the possibility of producing a spin-valve which only
allows electrons of one value of the spin to go through. We compare our
results for the conductance corrections with experimental observations.
We also propose an experimental test of our model of the contact regions.

\end{abstract}
\vskip .5 true cm

~PACS number: 72.10.-d, 85.30.Vw, 71.10.Pm, 73.40.Cg

\newpage

\section{\bf Introduction}

With the rapid advances made in the fabrication of high mobility
semiconductor heterojunctions, these systems have provided the 
setting for the discovery of several new phenomena in quantum 
systems. Popular examples include mesoscopic systems like 
quantum dots, quantum wires, and the two-dimensional electron gas samples 
in which the quantum Hall effects are observed \cite{datta}. 
In particular, quantum wires are created by the electrostatic gating of 
two-dimensional electron gases (2DEG) (with typical densities of 
$n_{2DEG} \sim 0.5 - 6 \times 10^{15} m^{-2}$) in the inversion layer of GaAs 
heterostructures. These GaAs samples typically have a very high mobility 
(typically, $\mu \sim 3 - 8 \times 10^{2} m^{2} V^{-1} s^{-1}$) because there 
is very little disorder in them; the mean free path of an electron in the 
2DEG is of the order of $\lambda_{MF} \sim 5 - 20\mu m$. 
This makes it possible to create ballistic channels a few microns in length 
for studying electron transport in such wires, especially 
at low temperatures when the 
thermal de Broglie wavelength of the electron is comparable to the channel 
length. Furthermore, since it is possible to maintain a low carrier 
concentration in these wires, it becomes possible for transport to take place 
through only a few channels or even a single channel.

Thus, several observations [2-12]
of the quantization of the conductance in electron 
transport through such channels have been made over the last two decades. More 
recently, several new ways have been found to produce such channels and this 
has led to even more precise experimental studies. This has brought into focus 
novel aspects of electron transport in such channels, not all of which 
are understood as yet. 

Let us first briefly review some of the recent
experimental findings in quantum wires.
The first striking observation is that of a number of flat plateaus in the dc 
conductance which are separated by steps of roughly the same value 
\cite{yacoby}
\beq
g = N{\tilde g}~\frac{2e^2}{h} ~,
\eeq
where $N$ denotes positive integers starting from one. The factor 
${\tilde g} < 1$ and is found to vary with the length of the quantum 
wire and the temperature; it has been seen to be as low as $0.75$. In fact, 
the plateaus tend towards $N(2e^2/h)$ as either the temperature is raised or 
the length of the quantum wire is shortened \cite{tarucha,yacoby}. This seems 
to imply a uniform renormalization of the plateau heights for each channel 
in the quantum wire as a function of wire length and temperature 
\cite{yacoby,facer,liang2,reilly}. Also, 
the flatness of the plateaus appears to indicate an insensitivity to the 
electron density in the channel. Furthermore, kinks are observed on the 
rise of the conductivity to some of the lowest plateaus. One such kink 
has been named the ``$0.7$ effect" \cite{thomas,liang,reilly}. These kinks are 
seen to wash away 
quickly with increasing temperature \cite{thomas,liang} and an external 
magnetic field placed in-plane and parallel to the channels \cite{liang}. Also, 
upon increasing such a magnetic field, a splitting of the conductance steps 
is observed together with an {\it odd-even} effect of the renormalization of 
the plateau heights, with the odd and even plateau heights being 
renormalized by smaller and larger amounts respectively. Finally, at very 
high magnetic field, another kink is seen to arise near the first 
spin-split plateau. 

Several of these experimental observations have found no satisfactory 
explanation till date. It is the purpose of this work to provide a consistent 
framework within which most of these observations of transport in quantum 
wires can be explained. We will rely upon several concepts and techniques that 
have become popular in the study of interacting mesoscopic systems. These 
include the concept of bosonization, effective actions and the renormalization 
group (RG) \cite{haldane,kane,furusaki1}. To be precise, we will employ 
these techniques and ideas to understand the low energy transport properties 
of ballistic electrons in a finite quantum wire attached to two semi-infinite 
Fermi leads \cite{safi,maslov,furusaki}, but with a difference; the contacts 
of the quantum wire with its leads will themselves be modeled as short 
quantum wires with junction barriers at either end. The properties of the 
contacts are unaffected by the 
external gate voltage which causes the formation of the discrete sub-bands in 
the quantum wire. The junction barriers will be 
modeled as localized $\delta$-functions to account for the back-scattering of 
electrons due to the imperfect coupling between the quantum wire and the 2DEG 
reservoirs; these barriers will renormalize the conductance as 
observed in the experiments. We will also study the effect of external electric 
and magnetic fields on this system. The properties of such a model for the 
quantum wire will be seen to account for several of the experimental 
observations mentioned above, as well as predict the possibility of some more 
interesting observations in future experiments in these systems. It should 
be stated here that a possible mechanism for some of the experimental 
observations \cite{yacoby} has been proposed in 
Ref. \cite{alekseev}; this is based on the 
anomalously enhanced back-scattering of electrons entering the 2DEG reservoir 
from the quantum wire due to the formation of Friedel oscillations of the 
electron density near the edges of the reservoirs, and it neglects 
interactions between the electrons in the quantum wire. Our model, however, 
attempts to understand these observations keeping in mind the importance 
of electron-electron interactions, barrier back-scattering, finite
temperature and magnetic field as well as all
length scales in the quantum wire system. 

The paper is organized as follows. In Sec. 2, we discuss the basics of the 
model outlined above. We show that the model can be described by a
$K_L$-$K_C$-$K_W$ Luttinger model
\cite{sumathi} with three different interaction parameters 
in the lead ($K_L$), the contacts ($K_C$) and the wire ($K_W$). By assuming
that the electron-electron interactions get screened out rapidly as one goes 
from one region to another, we get in a natural way the existence of 
localized barriers at the junctions. We then discuss how our model goes
beyond the concept of {\it ideal} contact resistances as studied in the 
Landauer-Buttiker formalism. In Sec. 3, we study the effective action of 
our model in the presence of external electric fields after integrating 
out all bosonic fields except those at the boundaries between the various 
regions - the leads, the contacts and the wire. Depending on the
relative sizes of the contacts and the wire, we define two regimes
- a) the quantum wire (QW) limit, where
the length of the wire is much greater than the contact, and b) the 
quantum point contact (QPC) limit, where the length of the contacts is 
much greater than the length of the wire. We then study the symmetries
of the effective action to determine when resonant transmission is possible 
as a function of a tunable gate voltage. All of this is first done
for spinless fermions and subsequently, we give the modifications of
the results for spinful fermions. In Sec. 4, we explicitly compute
the corrections to the conductance due to the barriers at finite
temperature ($T$) and for a finite length of the wire ($l$) and for finite 
contact lengths ($d$). We compute the frequency dependent Green's
functions of the model, in the different frequency regimes and use the Kubo 
formula to compute the conductance.
We also show how these results could have been anticipated from the 
renormalization group (RG) equations for the barrier strengths. In Sec. 5, 
we study our model in the presence of an external in-plane
magnetic field. Using RG 
methods, we show that the spin-up and spin-down electrons see different 
barrier heights at the junctions, and use this idea to explain the 
odd-even effect mentioned earlier. We also outline all the 
possible resonances that can be seen under such conditions. We point out
the possibility of producing a spin-valve at moderate magnetic fields.
In addition, we compute the conductance of our model 
and discuss its qualitative features as a function of the strength of
the magnetic field. In Sec. 6, we compare the features of 
the conductance expressions obtained with the observations made in 
various experiments for transport in quantum wires with and without an external 
magnetic field. We find that our model is applicable to a large class of 
experiments and gives a unified and qualitatively correct explanation of all 
of them. In particular, our model gives a possible explanation 
for the uniform renormalization of all the conductance steps seen in several
experiments. We also explain the odd-even effect seen in
experiments in the presence of magnetic field. In addition, we propose
more precise experimental tests of our model. 
Finally, we end in Sec. 7, with a summary of all the new results in our paper,
and outline further investigations that are possible. 

\section{\bf The Model}

In this section, we will study the Tomonaga-Luttinger liquid (TLL)
model \cite{haldane} of a quantum wire of finite length with no disorder, 
which is connected to the two 2DEG reservoirs modeled as two semi-infinite 
Fermi leads through two contact regions. The contacts are modeled
as short quantum wires with the junctions at either end modeled as 
$\delta$-function barriers. The inter-electron interactions in the
system, and hence the parameter $K$ which characterizes the interactions, vary 
abruptly at each of the junctions. Hence, we study a
$K_L$-$K_C$-$K_W$-$K_C$-$K_W$ model (see Fig. 1). The motivation for the 
above model is as follows. The electrons in the 2DEG are basically 
free, and hence, in the equivalent
1D model, they are modeled as semi-infinite leads with 
Luttinger parameter $K_L=1$. This can be understood as follows: if each end 
of the quantum wire is approximated by a point, only those 
electrons in the 2DEG which are in a zero angular momentum state (with respect 
to the appropriate end) can enter (or leave) the wire. Thus, the wave function 
of such a state has the radial coordinate as its only variable and we may, 
therefore, model the 2DEG as noninteracting 1D systems lying 
on either side of the quantum wire. The electron velocity in the leads 
$v_L$ is given by the Fermi velocity of the 2DEG electrons in the reservoirs 
$v_F = \sqrt{2E_{F2D}/m}$. On the other hand, the externally applied 
gate voltage $V_G$ is applied over a small region and this leads to the 
formation of several discrete sub-bands where the electrons feel the 
transverse confinement potential produced by $V_G$. This region is the 
one-dimensional quantum wire where the density of electrons is controlled by 
the gate voltage. The lowest energies $E_s$ in each sub-band are given by the 
discrete energy levels for the transverse confinement potential (and can 
therefore be shifted by changing $V_G$) \cite{buttiker}. The Fermi energy in 
the $s^{\rm th}$ channel is given by $E_{F1D} = E_{F2D} - E_s$. A channel is 
open when $E_{F1D} > 0$; the electron velocity $v_{W}(e)$ in the channel 
is then related to the 2DEG Fermi velocity $v_F$ by $v_W (e) = {\sqrt {v_F^2 - 
2 E_s /m}}$. In this gate voltage constricted region, the electrons will be 
considered as interacting via a short range (Coulomb-like) repulsion. 
Thus each discrete channel is modeled by a separate TLL. Let us, for the 
moment, consider one such channel with an interaction parameter $K_W$ and 
quasiparticle velocity $v_W$. 

The contacts represent the regions where the geometry changes from 
two-dimensional (2D) to 1D. In these regions of 
changing geometry, interactions between 
the electrons are likely to be very important; thus we model
the contact region as a Luttinger liquid with $K=K_C$. However, the gate 
voltage $V_G$ is unlikely to affect the properties of the electrons in these 
regions as the discrete sub-bands form a little deeper inside the wire. 
We choose different parameters $K_W$ for the wire and $K_C$ for the contact, 
because it is not obvious that inter-electrons interactions 
within the quantum wire will be the same as in the contact. The density of 
electrons in the quantum wire is controlled by the gate voltage, whereas the 
density of electrons in the contacts is controlled by the density of the 2DEG 
at or near the Fermi energy. Hence, we expect $K_C$ to be independent of 
$V_G$, but $K_W$ is dependent on $E_{F1D}$, which, in turn depends on $V_G$.
We will also show below that the change in the 
inter-electron interactions in each of the lead, contact and quantum wire 
regions gives rise to barrier-like back-scattering of the electrons. 

Simpler versions of this model (but without junction barriers and without 
contacts) have been studied by several authors \cite{safi,maslov,furusaki} 
who found perfect conductance through the TLL channel which is independent 
of the inter-electron interactions. Perfect conductance is also seen in 
several of the experiments \cite{wees,tarucha,yacoby,thomas,liang}. 
In the opposite limit, the model of a finite quantum wire connected to the 
two reservoirs by tunneling through very large barriers has also been 
studied \cite{fabrizio}. The idea of modeling 2DEG reservoirs by 1D 
noninteracting Fermi leads has also earlier been 
employed in studies of the fractional quantum Hall effect edge states coupled 
to Fermi liquids through a tunneling term in the Hamiltonian \cite{chamon}. 
Some studies of disordered quantum wires in such a model (again with perfect 
junctions) have also been conducted and the corrections to the conductance 
because of back-scattering impurities found \cite{safilong,maslovlong}. 
The continuity of 
the results found in these studies (which have quantum wires of a finite 
length) with those found earlier for infinite quantum wires \cite{kane} has 
also been established \cite{furusaki,safilong}. 

The main difference between our model and the earlier studies of the
quantum wire is that here we explicitly model the contacts as short
TLL wires bounded by junction barriers on either end and whose properties 
are unaffected by the gate voltage $V_G$. As we will discuss later, an 
experiment performed recently \cite{picciotto}
has conclusively shown the existence of a 
region (of an appreciable length of $2-6\mu m$) in between the quantum 
wire and the 2DEG reservoirs which leads to the back-scattering of 2DEG 
electrons entering the quantum wire. Furthermore, the idea that the 
properties of a one-dimensional system are determined by the Fermi 
energy of the 2DEG reservoirs has been used in Ref. \cite{matveev} to 
study the quantum point contact. In addition, we assume that the changes 
in the inter-electron interactions take place abruptly in going from the 
contacts into the quantum wire and that all inter-electron interactions get 
screened out very quickly in going from the contacts into the leads. It can, 
however, be shown that a smoother variation of the interaction parameter $K$ 
upon going from the quantum wire into the contacts and in going from the 
contacts into the noninteracting leads does not affect any of the transport 
properties in the $\omega \to 0$ (dc)
limit as long as we have no barriers of any kind in the 
system. We will now show that changes in the inter-electron 
interactions at the lead-contact and contact-quantum wire junctions give rise 
to barrier-like terms in the Hamiltonian of the system; the existence of these
terms is mentioned briefly in the work of Safi and Schulz \cite{safilong}. 
This is, however, only one reason why the junctions between the 
$1$D channel and its leads can cause the back-scattering of electrons; another 
reason is clearly the change in geometry 
in going from the 2DEG reservoirs into the 1D channel. 
This cause for the drop in the conductance of the channel has earlier been 
studied within the purview of the Landauer-Buttiker formalism; see
\cite{yacimry} and references therein. 

Let us begin by studying the simpler case of a quantum wire (in which electrons 
are interacting with each other) connected directly to the noninteracting, 
semi-infinite leads without any intermediate contact regions. Then there is 
only a single change in inter-electron interactions from zero in the leads to 
a finite value in the quantum wire. The kinetic part of the Hamiltonian for 
this system of interacting spinless electrons in the quantum wire when 
expressed in terms of the bosonic field $\phi(x)$ and its canonically 
conjugate momentum $\Pi(x)= \partial_t \phi/v_F$ is given by
\beq 
H_0 = \frac{1}{2} \int dx v_F [ \Pi(x)^2 + (\partial_x\phi(x))^2 ] ~,
\eeq
where $v_F$ is the Fermi velocity of the electrons in the channel. The part of 
the Hamiltonian which characterizes the short-ranged density-density 
interactions between the electrons in a 1D channel of length $l$ is given by
\beq
H_{int} = \int_0^l dx \int_0^l dy ~{\cal U}(x,y) \rho(x) \rho(y) ~,
\eeq
where ${\cal U}(x,y)$ characterizes the strength of the 
density-density interactions between the electrons, and $\rho(x)$ is the 
electronic density at the point $x$. Using a truncated form of the Haldane 
representation for the electronic density in terms of the bosonic field 
$\phi(x)$ \cite{haldane}, the density $\rho(x)$ is given by 
\beq
\rho(x) = \frac{1}{\sqpi}\partial_x\bar\phi(x) ~[c_0 + 
2c_1\cos(2\sqpi \bar\phi(x))]~,
\eeq
where $\bar \phi(x) = \phi(x) + k_Fx/\sqpi$, $c_0 = 1$, $c_1 = 
\Lambda /(2k_F)$, and $k_F$ is the Fermi wave vector. $\Lambda$ is the 
ultraviolet cutoff ($\Lambda < O(E_{F1D})$); 
it is the energy limit up to which the linearization of
the bands and hence bosonization is expected to be applicable.
If we now characterize the short range inter-electron interactions 
by ${\cal U}(x,y) = {\cal U}_0 \delta(x-y)$, 
then we can substitute the expressions for the density and the inter-electron 
interaction into the interaction term in the Hamiltonian. This gives us
\bea
H_{int} &=& {\cal U}_0 \int_0^l dx ~[\{\partial_x(\frac{c_0}{\sqpi}
\bar\phi)\}^2 ~-~2\partial_x(\frac{c_0}{\sqpi}\bar\phi)\partial_x (
\frac{c_1}{\sqpi} \sin (2\sqpi \bar\phi)) \nonum \\
& & ~~~~~~~~~~~~~~+~ \{\partial_x (\frac{c_1}{\sqpi}
\sin (2\sqpi\bar\phi) )\}^2] \nonum \\
&=& {\cal U}_0 \int_0^l dx ~[\frac{c_0^2}{\pi}(\partial_x\bar\phi)^2
~+~2\frac{c_0^2k_F}{\pi}\partial_x\phi~+~2\frac{c_0c_1k_F}{\sqpi}
\partial_x\sin (2\sqpi \bar\phi) \nonum \\
& & ~~~~~~~~~~~~+~2\frac{c_0c_1}{\sqpi}\partial_x\phi\partial_x
\sin (2\sqpi \bar\phi) ~+~2\frac{c_1^2}{\pi}(\partial_x\bar\phi)^2
(\cos(4\sqpi\bar\phi) +1)] \nonum \\
&=& {\cal U}_0 \int_0^l dx ~[(\frac{c_0^2+2c_1^2}{\pi})
(\partial_x\phi)^2
~+~2k_F(\frac{c_0^2+2c_1^2}{\pi})\partial_x\phi~+~2\frac{c_0c_1k_F}{\sqpi}
\partial_x\sin (2\sqpi\bar\phi) \nonum \\
& & ~~~~~~~~~~~~+~2\frac{c_0c_1}{\sqpi}\partial_x\phi\partial_x
\sin (2\sqpi\bar\phi) ~+~2\frac{c_1^2}{\pi}(\partial_x\bar\phi)^2
\cos(4\sqpi\bar\phi) ].
\eea
We can now simplify this expression by noting that several of the terms above 
contain rapidly oscillating factors of $\cos(k_Fx)$ or $\sin(k_Fx)$ which 
make those terms vanish upon performing the 
integration (unless we are at very specific fillings of the electron density).
Thus, we can ignore the fifth term straightaway. The first term can 
be added to a similar term in $H_0$ where it renormalizes the velocity and 
introduces an interaction parameter $K$. 
The second term is a chemical potential term and that too can be accounted for 
by shifting the field $\phi$ accordingly. The third term is clearly a boundary 
term, and it gives us two barrier like terms at $x=0$ and $x=l$, with
\bea
H_{barrier} &=& 2{\cal U}_0\frac{c_0c_1k_F}{\sqpi} \int_0^l dx ~
\partial_x \sin (2\sqpi\bar\phi) \nonum \\
&=& 2{\cal U}_0\frac{c_0c_1k_F}{\sqpi}~
[\sin (2\sqpi\phi(l)+ 2k_Fl)~-~\sin (2\sqpi\phi(0)) ].
\eea
Finally, the fourth term can also be rewritten as
\bea
H_{int,4} &=& 4{\cal U}_0\frac{c_0c_1}{\sqpi} \int_0^l dx ~(\partial_x
\phi)^2 \cos (2\sqpi\bar\phi)~+~2{\cal U}_0\frac{c_0c_1k_F}{\sqpi} \int_0^l dx 
~\partial_x\sin (2\sqpi\bar\phi) \nonum \\
& & -~4{\cal U}_0\frac{c_0c_1k_F^2}{\sqpi} \int_0^l dx ~\cos (2\sqpi\bar\phi)~,
\eea
in which the first and third terms again vanish because they contain rapidly 
oscillating factors within the integrals, and the second term adds on to 
$H_{barrier}$ exactly. All this finally gives us two $\delta$-function 
barriers at the junctions of the quantum wire with its Fermi liquid leads as 
\beq
H_{barrier} = 4{\cal U}_0 \frac{c_0c_1k_F}{\sqpi}~[ \sin(2\sqpi\phi(l)+ 
2k_Fl)~-~\sin (2\sqpi\phi(0))] ~.
\eeq
The extension of the derivation given above to our model with two 
intermediate contact regions where the inter-electron interactions are 
${\cal U}(x,y) = {\cal U}_1 \delta(x-y)$ 
(i.e., different from that in the quantum wire) is straightforward, 
and it yields four barrier terms: two at the junctions of the 
contacts with the leads, and two at the junctions of the contacts with the 
quantum wire. It is also very likely that the inner two barriers are much 
weaker than the outer two since the change in inter-electron interactions in 
going from the contacts to the quantum wire is likely to be much smaller than 
that in going from the contacts to the leads; also the change in geometry at 
the contact-quantum wire junction is likely to be much more adiabatic. Thus, 
we will from now on consider the junctions between the wire and the contacts 
and between the contacts and the leads as local barriers whose heights 
are determined by several factors, such as the nature of the inter-electron 
interaction and its screening, and the deviations from adiabaticity in the 
change in geometry in going from the reservoirs into the contacts or from the 
contacts into the quantum wire. To be general, we should take these four 
barriers to have different heights but it is very likely that any asymmetry 
between the left two and right two contacts will be small. Thus, we
can finally write the complete Hamiltonian for the quantum wire of 
spinless electrons, its contacts and its leads as 
\bea
H &=& (\int_{-\infty}^0+\int_{l+2d}^\infty ) ~
dx ~\frac{v_F}{2} [ \Pi(x)^2 + (\partial_x\phi(x))^2] \nonum \\ 
&& +(\int_0^d+\int_{l+d}^{l+2d}) ~
dx ~\frac{v_C}{2K_C} [\Pi_C(x)^2 + (\partial_x\phi(x))^2] \nonum \\
&& +\int_d^{l+d} ~dx ~\frac{v_W}{2K_W} [\Pi_W(x)^2 + 
(\partial_x\phi(x))^2] +V_{LC}\sin (2\sqpi\phi(0)) \nonum \\ 
&& +V_{CW}\sin(2\sqpi\phi(d)+ 2k_Fd) 
+V_{WC}\sin(2\sqpi\phi(l+d)+ 2k_F(l+d)) \nonum \\ 
&& +V_{CL}\sin(2\sqpi\phi(l+2d)+ 2k_F(l+2d))~,
\label{model}
\eea
where $\Pi_{C,W}(x) = (1/v_{C,W})\partial_t \phi(x)$.
Finally, it is worth commenting here that since our
model shows that a quantum wire with no disorder 
already has back-scattering junctions built into it, the notion of {\it ideal 
contact resistances} (which are 
seen in a study of this system using the Landauer-Buttiker formalism and arise 
from the {\it ideal} connection of the quantum wire to its reservoirs) which 
are universal in value, $h/2e^2$ to be precise, does not seem 
to hold true even for the so-called clean quantum wire with adiabatic 
junctions in the presence of inter-electron interactions within the quantum 
wire. We will show later that these junction barriers are likely to be weak 
when the lengths of the quantum wires are quite short or temperatures are
not very 
low, and that the junction barriers are likely to remain weak even after some 
small renormalization that might take place due to the electron-electron 
interactions in the quantum wire. Thus, the contact resistances between the 
wire and the reservoirs due to the junction barriers will be very nearly the 
universal value quoted above only for very short quantum wires (i.e., quantum 
point contacts) or when the 
temperatures are not very low. This is also observed in all the experiments 
till date \cite{tarucha,yacoby,thomas,liang}.

The generalization of the model to spinful fermions is straightforward. For 
completeness, we give below the Hamiltonian for spinful electrons in a quantum 
wire connected to external reservoirs through the contacts and junction 
barriers, 
\bea
H_{spin} &=& \sum_{i=\ua,\da}\Big[ (\int_{-\infty}^0+\int_{l+2d}^\infty) ~
dx ~\frac{v_{iF}}{2} [ \Pi_i(x)^2 + (\partial_x\phi_i(x))^2] \nonum \\ 
&& +(\int_0^d+\int_{l+d}^{l+2d}) ~dx ~\frac{v_{iC}}{2K_{iC}}[\Pi_{iC}
(x)^2 + (\partial_x\phi_i(x))^2] \nonum \\
&& + \int_d^{l+d} ~dx ~\frac{v_{iW}}{2K_{iW}} [\Pi_{iW}(x)^2 + 
(\partial_x\phi_i(x))^2] +V_{iLC}\sin (2\sqpi\phi_i(0)) \nonum \\ 
&& + V_{iCW}\sin(2\sqpi\phi_i(d)+ 2k_{iF}d) 
+V_{iWC}\sin(2\sqpi\phi_i(l+d)+ 2k_{iF}(l+d))\nonum \\ 
&& + V_{iCL}\sin(\sqpi\phi_i(l+2d)+ 2k_{iF}(l+2d)) \Big].
\label{modelspin}
\eea
Note that we have allowed for independent velocities and 
interaction strengths for the $\ua$ and $\da$ electrons. This
generality will be required when we study the model in the presence of
a magnetic field. Finally, let us note the fact that we will be taking into 
account only the outer two junction barriers (i.e., those at the junctions of 
the contacts and the leads) in all our subsequent calculations as these are 
likely to be the more significant junction barriers in the system as 
long as transport through fully open quantum wires is considered.

\section{Effective Actions}

In this section, the aim is to obtain an effective action 
in terms of the fields at the junction barriers for both spinless
and spinful electrons. We then analyze the symmetries of the 
effective action and obtain the resonance conditions.

\subsection{The case of spinless electrons}

In Sec. 2, it was shown that the screening out of the 
interactions in the 2DEG leads to a Hamiltonian with junction barriers given 
in Eq. (\ref{model}). The effective action for this model of spinless 
electrons can be written as
\beq
S = S_0 + S_{barrier} + S_{gate} ~,
\eeq
where we have defined each of the actions separately below.
\beq
S_0= \int d\tau ~[ \int_{-\infty}^0 dx {\cal L}_1 + \int_0^d dx {\cal
L}_2 + \int_d^{l+d} dx {\cal L}_3 + \int_{l+d}^{l+2d} dx {\cal L}_2
+\int_{l+2d}^{\infty} dx {\cal L}_1 ]~,
\label{s0}
\eeq
where
\beq
{\cal L}_1 ~=~ {\cal L} (\phi ; K_L , v_L ), ~~~ 
{\cal L}_2 ~=~ {\cal L} (\phi ; K_{C} , v_{C} ),
~~~{\rm and} ~~~ {\cal L}_3 ~=~ {\cal L} (\phi ; K_{W} , v_{W}) ~,
\label{sol} 
\eeq
and we have defined ${\cal L} (\phi ; K,v) = (1/2Kv) (\pt \phi)^2 +(v/2K) 
(\px \phi)^2$, and used the imaginary time $\tau=it$ notation.
\bea 
S_{barrier} &=& \int d\tau ~\Lambda [ V_1 \cos (2\sqpi\phi_1) + V_1
\cos (2\sqpi\phi_4 +2k_FL) \nonum \\ 
&& ~~~~~~~~~ +V_2\cos(2\sqpi\phi_2 + 2k_Fd) + V_2\cos(2\sqpi\phi_3 +
2k_F(l+d)) ]~,
\eea 
where we have set $V_{LC} = V_{CL}=V_1 \Lambda$ and 
$V_{CW}=V_{WC} =V_2 \Lambda$ assuming 
left-right symmetric barriers ($V_1$ and $V_2$ are dimensionless), and have 
used $\phi(0,\tau) = \phi_1(\tau) 
\equiv \phi_1$, $\phi(d,\tau) = \phi_2(\tau) \equiv \phi_2$, $\phi(l+d,\tau)
= \phi_3(\tau) \equiv \phi_3$ and $\phi(L,\tau) = \phi_4(\tau) \equiv
\phi_4$. The total length of the wire is denoted by $L=l+2d$. 
We shall henceforth assume that $V_2\ll V_1$ and can be
dropped; as we have explained earlier, the inner two
barriers are likely to be weaker than the outer two barriers. 
We also include the coupling of the electrons in the wire 
to an external gate voltage $V_G$ given by 
\beq
S_{gate} ~=~ V_G \int d\tau \int_d^{l+d} dx \rho(x,\tau) ~
=~ V_G\int d\tau {[\phi_3 - \phi_2]\over \sqpi}.
\eeq
This coupling is necessary because it is the
gate voltage which controls the density of electrons in the wire,
which, in turn, controls the number of channels in the quantum
wire. Experimentally, an external voltage drop across the wire drives
the current through the wire, which is measured as a function of
the gate voltage or the density of electrons in the wire. 

Since the Luttinger liquid action is quadratic, the effective action 
can be obtained in terms of the fields $\phi_i$, $i=1...4$, by integrating
out all degrees of freedom except those at the positions of the
four junction barriers, following Ref. \cite{kane}. 
Using the (imaginary time) Fourier transform of the fields 
\beq
\phi_1(\tau) = \sum_{\on} e^{-i\on \tau} \tphi_{1n} (\on),
\quad \phi_2(\tau) = \sum_{\on} e^{-i\on \tau} \tphi_{2n} (\on) ~,
\eeq	
we explicitly obtain the $S_0$ part of the effective action; this is
presented in Appendix A. (The $\on$ are the Matsubara frequencies which are
quantized in multiples of the temperature as $\on = 2\pi n k_B T$).
In the high frequency limit, or, equivalently at high temperatures,
where $\on d/v_C, \on l/v_W \gg 1$, the effective action reduces to 
\beq
S_{0,eff,high} (\tphi_1,\tphi_2,\tphi_3,\tphi_4) 
={K_L+K_C \over 2K_L K_C} \sum_{\on} |\on| (\tphi_1^2+\tphi_4^2)
+{K_W+K_C \over 2K_W K_C} \sum_{\on} |\on| (\tphi_2^2+\tphi_3^2).
\label{highfreq}
\eeq
In this limit, all the barriers are seen as the sum of individual 
barriers with no interference. In fact, if we integrate out the two
inner fields $\phi_2$ and $\phi_3$, we are just left with 
\beq
S_{0,eff,high}(\tphi_1,\tphi_4)
={K_L+K_C \over 2K_L K_C} \sum_{\on} |\on| (\tphi_1^2+\tphi_4^2).
\eeq
The surprising point to note is that the effective interaction
strength $K_{eff} =K_L K_C /(K_L+K_C)$ depends only on the
interaction strengths in the contacts and in the leads (where there are
no interactions), and not on the interaction strength in the wire! 
Furthermore, since the gate voltage $V_G$ couples only to the inner fields 
$\phi_2$ and $\phi_3$ and these two fields are completely decoupled 
from the outer fields $\phi_1$ and $\phi_4$ in ${\cal L}_{0,eff,high} 
(\phi_1,\phi_2,\phi_3,\phi_4)$ above, integrating out $\phi_2$ and 
$\phi_3$ does not lead to any gate voltage term in the final effective 
action in this temperature regime.

Depending on whether $d\gg l$ or $l\gg d$, we can have two possible
scenarios of intermediate regimes, each with two 
crossovers. We can express all our lengths in terms of equivalent 
temperatures by defining $v_C/ d = k_B T_d$ and $v_W/l = k_B T_l$. 
So the high temperature limit defined above is just $T\gg T_d,T_l$. 

\begin{itemize}

\item{}

Let us first consider the quantum wire limit where $l\gg d$.

In the intermediate frequency (or temperature) regime of 
$T_l \ll T \ll T_d$, the action becomes
\bea
S_{0,eff,int} (\tphi_1,\tphi_2,\tphi_3,\tphi_4) &=&
{1\over 2K_L} \sum_{\on} |\on| (\tphi_1^2+\tphi_4^2) +
{1\over 2K_W} \sum_{\on} |\on| (\tphi_2^2+\tphi_3^2) \nonum \\
& & + {U_C \over 2} \sum_{\on} [ (\tphi_1- \tphi_2) ^2 
+(\tphi_3- \tphi_4)^2] + S_{gate} ,
\eea
where $U_C = v_C/(K_Cd)$ is an energy whose significance will 
become clear shortly. As the action is quadratic, we can integrate 
out $\tphi_2$ and $\tphi_3$ to be left with an action dependent only 
on $\tphi_1$ and $\tphi_4$ as given by
\bea
& & S_{0,eff,int} (\tphi_1,\tphi_4) \nonum \\
&=& {1\over 2K_L} \sum_{\on} |\on| (\tphi_1^2+\tphi_4^2) \nonum \\
& & + ~~{1\over 2K_W} \sum_{\on} |\on| 
[\frac{U_C^2}{A^2}(\tphi_1^2+\tphi_4^2) + 
\frac{2{\tilde V}_G U_C}{A^2}(\tphi_1-\tphi_4) ] \nonum \\
& & + ~~\frac{U_C}{2} \sum_{\on} \Bigl[ [ (\frac{U_C}{A} - 1)\tphi_1 + 
\frac{{\tilde V}_G}{A}]^2 + [ (\frac{U_C}{A} - 1)\tphi_4 - 
\frac{{\tilde V}_G}{A}]^2 \Bigr] \nonum \\
& & + \int d\tau \frac{{\tilde V}_G U_C}{A}(\phi_4 - \phi_1) ~,
\eea
where $A = U_C + |\on|/K_W$ and ${\tilde V}_G = V_G/\sqrt{\pi}$. We can 
approximate $A$ by $U_C$ which is justified in the 
intermediate regime as $T \ll T_d$ and $K_C \sim K_W$. 
Then we are finally left with the expression
\beq
S_{0,eff,int} (\tphi_1,\tphi_4) = {{K_L+K_W}\over{2K_LK_W} } \sum_{\on} |\on| 
(\tphi_1^2+\tphi_4^2) ~+~ \int d\tau {\tilde V}_G(\phi_4 - \phi_1) ~.
\label{intertemp}
\eeq

\item{}

Now, we consider the QPC limit where $d \gg l$.

In the regime where $T_d \ll T \ll T_l$, the action becomes
\bea
S_{0,eff,int} (\tphi_1,\tphi_2,\tphi_3,\tphi_4) &=&
{1\over 2K_L} \sum_{\on} |\on| (\tphi_1^2+\tphi_4^2) \nonum \\
& & + {1\over 2K_C} \sum_{\on} |\on| (\tphi_1^2+\tphi_2^2+
\tphi_3^2+\tphi_4^2) \nonum \\
& & + {U_W \over 2}\sum_{\on} (\tphi_2- \tphi_3)^2 + S_{gate} , 
\eea
where $U_W = v_W/(K_Wl)$ is again a frequency independent
energy. As before, we can integrate out $\tphi_2$ and $\tphi_3$ to be left 
with an action dependent only on $\tphi_1$ and $\tphi_4$ given by
\beq
S_{0,eff,int} (\tphi_1 ,\tphi_4 ) ~=~ S_{0,eff,high} (\tphi_1 ,\tphi_4 )~. 
\eeq
Thus there is no difference between the intermediate and high energy scales in
the QPC limit because the gate voltage is applied over too short a length to
affect the conductance even at intermediate temperatures.

\end{itemize}

Finally, in the low frequency limit where $\on \ll v_W/l , v_c/d$ 
(i.e., $T \ll T_d$ and $T \ll T_l$), $S_0$ reduces to 
\bea
S_{ 0,eff,low} (\tphi_1,\tphi_2,\tphi_3,\tphi_4) 
&=& {1\over 2K_L} \sum_{\on} |\on| (\tphi_1^2+\tphi_4^2) + {U_C\over 2} 
\sum_{\on} [ (\tphi_1- \tphi_2) ^2 +(\tphi_3- \tphi_4)^2] \nonum \\
&& ~~~+{U_W\over 2} \sum_{\on} (\tphi_2- \tphi_3)^2 ~.
\eea
Since the action is still quadratic, it is possible to integrate out
the two inner fields $\tphi_2$ and $\tphi_3$ and get the effective action
wholly in terms of the $\tphi_1$ and $\tphi_4$ fields, remembering
however, to also include the gate voltage term which couples to the inner 
fields. After doing out, we are left with the full effective action as
\bea
S_{eff,low} (\phi_1,\phi_4) &=& S_{0,eff,low} + 
S_{gate}+ S_{barrier} \nonum \\
&=& {1\over 2K_L} \sum_{\on} |\on| (\tphi_1^2+\tphi_4^2) +
{U_C U_W \over 2(U_C+2U_W)}\sum_{\on} (\tphi_1-\tphi_4
-{{\tilde V}_G \over U_W})^2. \nonum \\
&&
\eea 
In this limit, the full action can be rewritten in terms of a 
``current" field $\chi(\tau)$ and a ``charge" field $n(\tau)$
(and their Fourier transforms $\tilde \chi$ and $\tilde n$) defined as 
\beq
\chi(\tau) = {\phi_1 +\phi_4 \over 2 }+ {k_F L\over 2\sqpi}, \quad {\rm 
and} \quad n(\tau) = {\phi_4 - \phi_1 \over \sqpi} + {k_F L\over \pi} ~.
\eeq
The action is given by 
\bea
S_{eff,low} &=& S_0 + S_{barrier} +S_{gate} \nonum \\
&=& {1\over 2K_L} \sum_{\on} |\on| [(\tchi - {k_F 
L\over 2\sqpi})^2 + {\pi\over 4} (\tilde n -{k_F L\over \pi})^2] \nonum \\
&& + \int d\tau [{U_{eff} \over 2} (n-n_0)^2 + V_1 \Lambda 
\cos (2\sqpi \chi) \cos (\pi n) ] ~,
\label{lowfreq}
\eea
where $n_0 = (2k_C d + k_W l)/\pi - V_G/(\pi^{3/2}U_W)$ and 
$U_{eff} =\pi U_C U_W / (U_C+2U_W)$. The derivation of the effective 
action in this limit follows the method outlined 
in Ref. \cite{kane}; however, their derivation was for a uniform wire with a 
single interaction parameter $K$, whereas we have three interaction 
parameters here. $K_W$ acts only within the quantum wire delimited by the two
contact regions, $K_C$ acts within the contact region, and $K_L =1$ outside 
the contact and wire region. 
The current field is interpreted as the number of particles transferred 
across the two barriers, and the charge field is the number of particles 
between the barriers. In the low frequency limit, the two barriers are 
clearly being seen as one coherent object with charge and current degrees of 
freedom. Since
in the limit of weak barriers, $V_1 \ll U_{eff}$, the action is minimized when 
$n=n_0$, we can integrate out the quadratic fluctuations of $n-n_0$ to obtain 
an effective action only in terms of the single variable $\chi$; we obtain
\beq
S_{low} = \int d\tau ~[~2V_1 \Lambda \cos (2\sqpi\chi) \cos (\pi n_0) 
-{2(V_1 \Lambda)^2 \over U_{eff}} (\pi \cos (2\sqpi \chi) \sin (\pi n_0))^2 
+ \cdots ~]~.
\eeq
The first term in this effective action is precisely the
same term that is obtained for the impurity potential for 
a single barrier in terms of the variable $\chi$.

In the low frequency limit, from Eq. (\ref{lowfreq}), we see that the 
effective action contains extra terms due to the interference between the 
two barriers. It is easy to check that this effective action is invariant
under $\chi\rightarrow \chi+\sqrt{\pi}, n \rightarrow n$; this is 
the same symmetry which exists for a single barrier \cite{kane}, and it
corresponds to the transfer of a single electron across the two barriers, and 
hence in our model, from the left lead to the right lead. But when $n_0$ is 
precisely equal to a half-odd-integer, the action is also invariant 
under $\chi \rightarrow \chi+\sqpi/2, n\rightarrow 2n_0-n$. As explained in
Ref. \cite{kane}, this corresponds to the `transfer of half an electron
across the wire' accompanied by a change in the charge state of the 
wire. In the language of scattering, this corresponds to resonant tunneling
through a virtual state. Within the TLL theory, this is the explanation
of the Coulomb blockade phenomenon, which leads to steps or plateaus
in the current versus gate voltage for quantum dots.

\subsection{The case of spinful electrons}

The spinless electron model is expected to be valid for real systems
in the presence of strong magnetic fields which completely
polarizes all the electrons in a given channel. 
However, for a real system without magnetic 
field, or in the presence of weak magnetic fields which do not
polarize all the electrons, one has to study a model of electrons
with spin. We shall study such a model here. Its modification due
to the presence of a magnetic field will be studied in Sec. 5.
The basic action of the model is a straightforward extension of
the model for spinless fermions given in Eqs. (\ref{s0}) and (\ref{sol}), 
with $\phi_{\ua}$ denoting the spin up boson and $\phi_{\da}$ denoting
the spin down boson. However, since the Coulomb interaction couples the spin 
up and spin down fermions (for instance, remember the Hubbard term which is 
$U\sum_i n_{i\ua} n_{i\da}$), the Luttinger model is diagonal only in terms of 
the charge and spin fields $\phi_\rho =(\phi_{\ua} + \phi_{\da})/{\sqrt 2}$ 
and $\phi_\sigma =(\phi_{\ua} - \phi_{\da})/{\sqrt 2}$. In terms of 
these fields $S_0$ is given by
\beq
S_0 = \int dt ~[ \int_{-\infty}^0 dx {\cal L}_1 
+ \int_0^d dx {\cal L}_2 + \int_d^{l+d} dx {\cal L}_3
+ \int_{l+d}^{l+2d} dx {\cal L}_2 + \int_{l+2d}^{\infty} dx {\cal L}_1 ]~, 
\label{s0spin}
\eeq
where
\bea
{\cal L}_1 ~&=&~ {\cal L} (\phi_\rho ; K_{L\rho} , v_L ) ~+~ 
{\cal L} (\phi_\sigma ; K_{L\sigma} , v_L ) ~, \nonum \\
{\cal L}_2 ~&=&~ {\cal L} (\phi_\rho ; K_{C\rho} , v_{C\rho} ) ~+~ 
{\cal L} (\phi_\sigma ; K_{C\sigma} , v_{C\sigma} ) ~, \nonum \\
{\cal L}_3 ~&=&~ {\cal L} (\phi_\rho ; K_{W\rho} , v_{W\rho} ) ~
+~ {\cal L} (\phi_\sigma ; K_{W\sigma} , v_{W\sigma} ) ~,
\eea
with $\cal L$ defined as before, ${\cal L} (\phi ; K,v) = (1/2Kv) 
(\pt \phi)^2 +(v/2K) (\px \phi)^2$,
$K_{L\rho} = K_{L\sigma}=1$ are the interaction parameters in
the two external leads, and $K_{C/W\rho}$ and $K_{C/W\sigma}$ are the
interaction parameters in the contacts and wire respectively.
As for the spinless case, we include junction barrier terms of the form
\beq
S_{barrier} = \int d\tau \sum_{i=\da,\ua} ~V_i \Lambda ~[\cos (2\sqpi\phi_i(0,
\tau) ) +\cos (2\sqpi\phi_i(L,\tau) +2k_F L)] ,
\eeq
at the junctions of the contacts and the leads, and we assume that the
barriers at the junctions of the wire and the contact are weak and can
be ignored. The barrier action can be re-expressed in terms of the 
diagonal fields of the model (using $V_{\ua} = V_{\da} =V$) as 
\beq
S_{barrier} = V \Lambda~\int d\tau [\cos (\sqrt{2\pi}\phi_{1\rho}) \cos (
\sqrt{2\pi} \phi_{1\sigma}) + \cos (\sqrt{2\pi} \phi_{4\rho} +2k_F L)\cos (
\sqrt{2\pi} \phi_{4\sigma})] ~,
\eeq
where, as before, we define $\phi_\rho(0) = \phi_{1\rho}$ and $\phi_\rho(L) = 
\phi_{4\rho}$ and similarly for the $\phi_\sigma$ fields. The gate voltage 
only couples to the charge degree of freedom within the wire region as
\beq
S_{gate} = ~\frac{V_G}{\sqpi} ~\int d\tau 
(\phi_{3\rho} - \phi_{2\rho}) ~, 
\eeq
where $\phi_\rho(d)=\phi_{2\rho}$ and $\phi_\rho(l+d)=\phi_{3\rho}$ as 
before. So just as in the spinless case, we can integrate out all degrees 
of freedom except at $x=0,d,l$ and $L$ and obtain the effective
action. The full details of the effective action are spelt out in
Appendix B. By taking its high, intermediate and 
low frequency limits, we will be able to 
obtain conductance corrections just as we did for the spinless fermions.

In the high frequency limit where $\omega \gg v_{Ca} /d$ and $v_{Wa} /l$, the
two barriers are seen as decoupled barriers, with
\bea
&& S_{0,eff,high} (\phi_{1,\rho/\sigma},\phi_{2,\rho/\sigma},
\phi_{3,\rho/\sigma},\phi_{4,\rho/\sigma}) \nonum \\
&& \quad = \sum_{\on}
\vert \on \vert [ \frac{K_{L\rho} + K_{C\rho}}{2 K_{L\rho} K_{C\rho}} 
(\tphi_{1\rho}^2 + \tphi_{4 \rho}^2 ) + \frac{K_{C\rho} + K_{W\rho}}{2 
K_{C\rho} K_{W\rho}} (\tphi_{2\rho}^2 ~+~ \tphi_{3\rho}^2 ) \nonum \\
&& ~~~~~~~~~~~~~ +{\rm ~~~similar ~terms ~with~} 
\rho \rightarrow \sigma ].
\eea
The fields $\phi_{2a}$ and $\phi_{3a}$ are completely decoupled
from the fields at $x=0$ and $L$ and can be integrated out yielding 
\bea
&& S_{0,eff,high} (\phi_{1,\rho/\sigma},\phi_{4,\rho/\sigma}) 
\nonum \\
&& \quad = {K_{L\rho}+K_{C\rho} \over 2K_{L\rho} K_{C\rho}} \sum_{\on} |\on| 
(\tphi_{1\rho}^2+\tphi_{4\rho}^2 ) + {K_{L\sigma}+ K_{C\sigma} \over 2 K_{L
\sigma} K_{C\sigma}} \sum_{\on} |\on| (\tphi_{1\sigma}^2+\tphi_{4\sigma}^2).
\eea
Just as for the spinless case, we see that the parameters of the wire
do not enter $K_{eff,\rho/\sigma} = K_{L\rho/\sigma} 
K_{C\rho/\sigma}/(K_{L\rho/\sigma}+K_{C\rho/\sigma})$. Nor does the
gate voltage affect the action. 

Just as in the spinless case, we have two possibilities for the intermediate
frequency regime, the QW limit or the QPC limit.

For the QW limit, we have $v_W/l \ll \on \ll v_C/d$, and we obtain 
\bea
&&S_{eff,int} (\tphi_{1,\rho/\sigma},\tphi_{2,\rho/\sigma},\tphi_{3,
\rho/\sigma}, \tphi_{4,\rho/\sigma}) \nonum \\ 
&& \quad = {1\over 2K_{L\rho}} \sum_{\on} |\on| (\tphi_{1\rho}^2+ 
\tphi_{4\rho}^2) + {1\over 2K_{L\sigma}} \sum_{\on} |\on| (\tphi_{1\sigma}^2 
+ \tphi_{4\sigma}^2) \nonum \\
&& \quad + {1\over 2K_{W\rho}} \sum_{\on} |\on| (\tphi_{2\rho}^2+
\tphi_{3\rho}^2) + {1\over 2K_{W\sigma}} \sum_{\on} |\on| (\tphi_{2\sigma}^2 + 
\tphi_{3\sigma}^2) \nonum \\ 
&& \quad + {U_{C\rho} \over 2} \sum_{\on} [ (\tphi_{1\rho}- \tphi_{2\rho}) ^2 
+ (\tphi_{3\rho}- \tphi_{4\rho})^2] \nonum \\
&& \quad + {U_{C\sigma} \over 2} \sum_{\on} [ (\tphi_{1\sigma}- 
\tphi_{2\sigma})^2 + (\tphi_{3\sigma}- \tphi_{4\sigma})^2] + S_{gate} ,
\eea
where $U_{C\rho,\sigma} = v_{C\rho,\sigma}/(K_{C\rho,\sigma}d)$ is the 
charging energy 
for the charge degrees of freedom. As the action is quadratic, we can 
integrate out the $\tphi_{2,\rho/\sigma}$ and $\tphi_{3,\rho/\sigma}$ spin 
and charge fields to be 
left with an action dependent only on the $\tphi_{1,\rho/\sigma}$ and 
$\tphi_{4,\rho/\sigma}$ spin and charge fields as given by
\bea
\hspace*{-1cm}S_{0,eff,int} (\phi_{1,\rho/\sigma}, \phi_{4,\rho/\sigma})
\hspace*{-0.3cm} &=& \frac{K_{L\rho}+K_{W\rho}}{2K_{L\rho}K_{W\rho}} 
\sum_{\on} |\on| (\tphi_{1\rho}^2+\tphi_{4\rho}^2) \nonum \\
&& \hspace*{-0.5cm}+ \frac{K_{L\sigma}+K_{W\sigma}}{2K_{L\sigma}K_{W\sigma}} 
\sum_{\on} |\on| (\tphi_{1\sigma}^2+\tphi_{4\sigma}^2) + \int d\tau 
{\tilde V}_G (\phi_{4\rho} - \phi_{1\rho} ) ,
\eea
where we have approximated
$U_C + \on/(K_{W,\rho/\sigma})$ by $U_C$; this is justified in the 
intermediate regime as $T \ll v_{C,\rho/\sigma}/d$ and 
$K_{C,\rho/\sigma} \sim K_{W,\rho/\sigma}$. 

In the QPC limit, we have
\bea
&&S_{eff,int} (\tphi_{1,\rho/\sigma},\tphi_{2,\rho/\sigma},\tphi_{3,
\rho/\sigma}, \tphi_{4,\rho/\sigma}) \nonum \\ 
&& \quad = {1\over 2K_{L\rho}} \sum_{\on} |\on| (\tphi_{1\rho}^2+
\tphi_{4\rho}^2) + {1\over 2K_{L\sigma}} \sum_{\on} |\on| (\tphi_{1\sigma}^2 
+ \tphi_{4\sigma}^2) \nonum \\
&& \quad + {1\over 2K_{C\rho}} \sum_{\on} |\on| 
(\tphi_{1\rho}^2 + \tphi_{2\rho}^2+\tphi_{3\rho}^2 +\tphi_{4\rho}^2)
+ {1\over 2K_{C\sigma}} \sum_{\on} |\on| (\tphi_{1\sigma}^2 + 
\tphi_{2\sigma}^2 + \tphi_{3\sigma}^2 +\tphi_{4\sigma}^2) \nonum \\ 
&& \quad + {U_{W\rho} \over 2}\sum_{\on} (\tphi_{2\rho}- \tphi_{3\rho}) ^2 
+ S_{gate} ,
\eea
where $U_{W\rho} = v_{W\rho}/(K_{W\rho}l)$ is the charging energy 
for the charge degrees of freedom in the wire. 
As before, we may integrate out the inner degrees of freedom to find that
\beq
S_{0,eff,int} (\tphi_{1,\rho/\sigma}, \tphi_{4,\rho/\sigma}) ~=~ 
S_{0,eff,high} (\tphi_{1,\rho/\sigma}, \tphi_{4,\rho/\sigma})
\eeq
as expected.

Finally, in the low frequency limit $\on \ll v_C /d$ and $v_W /l$, 
as in the spinless case, the terms multiplying $1/K_{C\rho/\sigma}$ and 
$1/K_{W\rho/\sigma}$ in Eq. (\ref{spineff}) in Appendix B 
become constant `mass' terms. We get the full effective action as
\bea
&&S_{eff,low}
(\phi_{1,\rho/\sigma},\phi_{2,\rho/\sigma},\phi_{3,\rho/\sigma},
\phi_{4,\rho/\sigma}) \nonum \\ 
&& \quad = \frac{1}{2 K_{L\rho}} \sum_{\on} |\on|
(\tphi_{1\rho}^2 + \tphi_{4\rho}^2) 
+ \int d\tau [\frac{v_{C\rho}}{K_{C\rho}d} \{ (\phi_{1\rho} -\phi_{2
\rho})^2 + (\phi_{3\rho} - \phi_{4\rho})^2 \} \nonum \\
&& \quad ~~ + \frac{v_{W\rho}}{K_{W\rho} } (\phi_{2\rho} - \phi_{3\rho})^2 
+ {\rm similar ~terms ~with ~} ~\rho \rightarrow \sigma ] 
+ \frac{eV_G}{\sqpi} \int d\tau [\phi_{3\rho} - \phi_{2\rho} ] \nonum \\
&& \quad ~~ + V \Lambda \int d\tau [\cos (\sqrt{2\pi}\phi_{1\rho}) \cos (
\sqrt{2\pi} \phi_{1\sigma}) + \cos (\sqrt{2\pi}\phi_{4\rho} +2k_F L)\cos (
\sqrt{2\pi} \phi_{4\sigma})] ~. \nonum \\
&&
\label{lowfreqspin}
\eea
Just as we did in the spinless case, we now integrate out the fields 
at $x=d$ and $l+d$, in terms of which the above action is quadratic, to get 
\bea
&&S_{eff,low} (\phi_{1,\rho/\sigma},\phi_{4,\rho/\sigma})
\nonum \\
&& \quad = {1\over 2K_{L\rho}}\sum_{\on} |\on| (\tphi_{1\rho}^2 +\tphi_{4
\rho}^2) + {U_{C\rho} U_{W\rho} \over 2(U_{C\rho}+2U_W{\rho})} \sum_{\on} 
(\tphi_{1\rho}-\tphi_{4\rho} -{{\tilde V}_G\over U_{W\rho}})^2 \nonum \\
&& \quad ~~ + {1\over 2K_{L\sigma}}\sum_{\on} |\on| (\tphi_{1\sigma}^2 +
\tphi_{4\sigma}^2) + {U_{C\sigma} U_{W\sigma} \over 2(U_{C\sigma}+2
U_W{\sigma})} \sum_{\on} (\tphi_{1\sigma}-\tphi_{4\sigma})^2. 
\eea 
Here, we see that the effective mass terms are given by $U_{eff,\rho}
=\pi U_{C\rho} U_{W\rho} / (U_{C\rho}+2U_W{\rho})$ and
$U_{eff,\sigma} =\pi U_{C\sigma} U_{W\sigma} / (U_{C\sigma}+2U_{W\sigma})$ for 
the `charge' and `spin charge' fluctuations respectively. 
We denote the `charge on the quantum wire' 
fields as $n_\rho=\sqrt{2/\pi} (\phi_{1\rho} - \phi_{4\rho})$ and $n_\sigma=
\sqrt{2/\pi} (\phi_{1\sigma} - \phi_{4\sigma})$ respectively, and 
the `current' fields as $\chi_\rho = (\phi_{1\rho} + \phi_{4\rho})/\sqrt{2}, 
\chi_\sigma= (\phi_{1\sigma} + \phi_{4\sigma})/\sqrt{2}$ along with their 
appropriate Fourier transforms $\tchi_{\rho/\sigma}$ and 
${\tilde n}_{\rho/\sigma}$ just as we did in the spinless case. The action 
then takes the following form,
\bea 
S_{eff,low} &=& {1\over 2K_{L\rho}} 
\sum_{\on} |\on| [|\tchi_{\rho} - {k_F d\over \sqpi}|^2 + {\pi\over 4} 
|\tilde n_{\rho} -{2k_F d\over \pi}|^2] + \int d\tau {U_{eff,\rho} \over 2} 
(n_{\rho} -n_{0\rho})^2 \nonum \\
&& ~+ {1\over 2K_{L\sigma}} \sum_{\on}|\on| [|\tchi_{\sigma}|^2 + {\pi\over 4} 
|\tilde n_{\sigma}|^2] + \int d\tau {U_{eff,\sigma} \over 2} 
(n_{\sigma})^2 \nonum \\
&& ~+ 2V \Lambda \int d\tau [\cos (\sqpi\chi_{\rho}) 
\cos ({\pi n_\rho\over 2}) \cos (
\sqpi\chi_{\sigma}) \cos({\pi n_\sigma\over 2}) \nonum \\ 
&& ~~~~~~~~~~~~~~~~ +\sin (\sqpi\chi_{\rho}) \sin ({\pi n_\rho\over 2}) \sin (
\sqpi\chi_{\sigma}) \sin ({\pi n_\sigma\over 2}) ]~.
\label{reson}
\eea
We have used the fact that since it is only the $\rho$ field which couples
to the gate voltage and not the $\sigma$ fields, we only get 
$n_{0\rho} = (2k_C d + k_W l)/\pi - V_G/(\pi^{3/2}U_{W\rho})$ and 
$n_{0\sigma} =0$.

We now study the symmetries of the effective action to find out the 
possible resonances. As in the spinless fermion case, this effective action 
is invariant under the transformation $\chi_{\rho} \rightarrow \chi_{\rho} + 
\sqpi$ and $\chi_{\sigma} \rightarrow \chi_{\sigma} + \sqpi$, which
corresponds to the transfer of either an up electron or a down
electron through the two barriers. But besides this symmetry, there
are also some special gate voltages at which one can get resonance 
symmetries. This can happen when we adjust the gate voltage so as to make 
$n_{0\rho}$ an odd integer. In that case, $V_{eff}$ is invariant under
\bea
&& n_\sigma \rightarrow -n_\sigma, ~~ n_\rho \rightarrow 2n_{0\rho} - n_\rho 
{\rm ~~in ~conjunction ~with} \nonum \\
&& {\rm either} ~~ (i)~\chi_\rho\rightarrow \chi_{\rho} +\sqpi, ~~
\chi_\sigma \rightarrow \chi_\sigma ~~ {\rm or} ~~ (ii)~\chi_\sigma 
\rightarrow \chi_{\sigma} +\sqpi, ~~ \chi_\rho \rightarrow \chi_\rho ~. 
\eea
As explained in Ref. \cite{kane}, this resonance which occurs when $n_{0\rho}$ 
is tuned to be an odd integer, is called a Kondo resonance because it happens 
when two spin states in the island with $n_{\sigma}=\pm 1$ become degenerate.

The kind of resonance which was seen for spinless fermions when two
charge states on the island becomes degenerate is harder to see for
spinful fermions. Two charge states become degenerate when $n_{0\rho}$
is tuned to be a half-odd-integer. But, in that case, the effective
action in Eq. (\ref{reson}) does not have any extra `resonance symmetry' 
unless $n_{0\sigma}$ (which we have set to be zero) is also tuned to be 
a half-odd-integer. But non-zero $n_{0\sigma}$ is only possible 
when there is an effective magnetic field or $SU(2)$ breaking field
just over the quantum wire. This is because the Zeeman term is given by the
Hamiltonian density
\beq
{\cal H}_{Zeeman} = -h(\partial_x\phi_\ua - \partial_x\phi_\da)
= -\sqrt{2} h\partial_x\phi_\sigma ~,
\eeq
and it does not lead to any boundary terms as long as the magnetic field
is felt through the full sample. However, 
although in current experiments it is not possible to tune the $SU(2)$
breaking to occur only between the two barriers, it could
be possible in future experiments. Hence it is of interest to look
for possible resonances in this case as well. We see that if one could arrange
to tune both the gate voltage and the magnetic field (adjusted to be
just over the quantum wire) so that $n_{0\rho}$ and $n_{0\sigma}$ are
both half-odd-integers, the effective action in Eq. (\ref{reson}) is symmetric
under $n_\rho \rightarrow 2n_{0\rho}-n_\rho$, $n_\sigma \rightarrow 
2n_{0\sigma}-n_\sigma$, $\chi_{\rho} \rightarrow \chi_{\rho} +\sqpi /2$, and
$\chi_{\sigma} \rightarrow \chi_{\sigma} +\sqpi /2$. 
This resonance is exactly analogous to the resonance that existed for
spinless fermions and corresponds to hopping an electron from either
of the leads to the wire. But since this requires the tuning of two
parameters, it is a `higher' order resonance and will be more
difficult to achieve experimentally.
 
In fact, if we allow for non-zero $n_{0\sigma}$, then the effective
action also has the symmetry
\bea
&& n_\rho \rightarrow -n_\rho, ~~ n_\sigma \rightarrow 2n_{0\sigma} - 
n_\sigma ~~ {\rm in ~conjunction ~with} \nonum \\
&& {\rm either} ~~ (i) ~\chi_\rho\rightarrow \chi_{\rho} +\sqpi, ~~ 
\chi_\sigma \rightarrow \chi_\sigma ~{\rm or~} ~~ (ii) ~\chi_\sigma 
\rightarrow \chi_{\sigma} +\sqpi, ~~ \chi_\rho \rightarrow \chi_\rho ~,
\eea
when $n_{0\sigma}$ is an odd integer and $n_{0\rho}=0$. But this is hard
to achieve, because one needs to tune the external gate voltage so as to 
cancel the field due to the presence of all the other electrons within the 
two barriers as well. Hence, this resonance will not be easy to see in 
experiments. Moreover, it will show up in the spin conductance and not the 
charge conductance.

In conclusion, we have studied in this section the effective actions
of our model for both spinless and spinful fermions, and used them to
obtain conductance corrections away from resonances (where the conduction
is perfect) as a function of finite temperature and finite length of
the wire. The same technique will again be used in Sec. 5, where it
will be used to study the symmetries and obtain the conductance
corrections of the quantum wire in the presence of a magnetic field.

\section{\bf Computation of the conductances}

In this section, we compute the conductances of our TLL quantum wire with 
contacts, two semi-infinite Fermi liquid leads and two weak barriers at the 
junctions of the contacts and the leads, for both spinless
and spinful electrons, perturbatively in the barrier strength. We explicitly 
derive an expression for the conductance to lowest order in 
barrier strength (quadratic) in terms of the Green's functions of 
the model. The RG flow of the barrier strengths has been incorporated 
through a function $\chi(x,y)$. Thus, the behavior of
the Green's functions in the different frequency regimes 
determines the conductance corrections. The conductance corrections for a 
simpler version of the model of the quantum wire (i.e., one in which the 
quantum wire is directly connected to the Fermi leads through two weak 
junction barriers) has already been studied by Safi
and Schulz \cite{safilong}, who used a real time formulation
and computed time-dependent Green's functions. The perturbative corrections 
in the Kane-Fisher imaginary time formalism was also extended to the case of 
finite length wires by Maslov \cite{maslovlong} and Furusaki and Nagaosa 
\cite{furusaki}, who computed frequency dependent Green's functions. For our 
model, with five distinct spatial regions with their boundaries, the real 
time picture of TLL quasiparticle waves reflecting back and forth
between the boundaries (as developed by Safi and Schulz \cite{safi}) is more 
cumbersome; hence, we use the imaginary
time formulation and compute frequency dependent Green's functions. 

\subsection{\bf The formulation of the conductance expressions}

The current through a clean quantum wire through which spinless 
electrons are traveling can be found using the Kubo formula
\beq 
j(x) = \lim_{\omega\rightarrow 0}\int dy \sigma(x,y,\omega) E(y,\omega) ~,
\eeq
where $\sigma(x,y,\omega)$ is the non-local conductivity and is related to the 
two-point Green's function $G(x,y,\omega)$ at finite frequency $\omega$ as
\beq
\sigma(x,y,\omega) = ~-i\frac{2 \omega e^2}{h} G(x,y,\omega) ~.
\label{cond}
\eeq
For our model of the quantum wire, $G_{\bo}(x,y)$ has been computed in 
Appendix D. Note that the real frequency $\omega$ is related to $\bar \omega$
(used in the earlier sections) by the analytic continuation $\omega = i
{\bar\omega} +\epsilon$. From Appendix D, we find that 
$G_\bo(x,y) = K_L/(2|\bo|)$ + non-singular terms in $\bo$ in the limit 
$\omega \rightarrow 0$ for our model. hence the dc conductance $g_0$ is given
by
\beq
g_0 = \lim_{\omega\rightarrow 0}\sigma(x,y,\omega) = \frac{e^2}{h}~.
\eeq
This shows perfect dc conductance through the system 
as in the earlier models without contacts\cite{safi,maslov}. 
This result remains unchanged for the case of electrons with spin, 
except for a multiplication of the conductance by a factor of two.

For a quantum wire in the presence of stationary impurities, 
an explicit expression for the conductance can be derived to lowest
(quadratic) order in the impurity strength from the partition function,
using perturbation theory \cite{kane,safilong,maslovlong}. The 
renormalization group (RG) equations for the barriers (discussed in detail in 
subsection 4.4) imply that the barrier strengths grow under renormalization.
However, it is only for very low temperatures or very long wire
lengths that there will be considerable renormalization. 
In real experimental setups, the length of the
wire is in the range of micrometers and the temperatures in the
range of a Kelvin; hence one does not expect much renormalization. 
Hence, it is expected that the barrier strengths remain small 
enough for perturbation theory to be applicable. We follow the methods 
of Safi and Schulz \cite{safilong} and Maslov \cite{maslovlong}, who 
derived explicitly a conductance expression for a non-translationally 
invariant system and obtained 
\beq
g = g_0 K_L ( 1 - {\cal R} )~,
\eeq
where {\cal R} is the perturbative correction to second 
order in the impurity strength. $\cal R$ is given by 
\beq
{\cal R} = g_0 K_L \sum_{m=1}^{\infty} m^2 c_m^2 {\cal R}^{(m)}~,
\eeq
where the $c_m$'s are the coefficients for the terms in the Haldane 
representation of the fermionic density, and ${\cal R}^{(m)}$ is the 
correction due to the back-scattering of {\it m} electrons given by
\beq
{\cal R}^{(m)} = \int\int dx dy V(x)V(y) \cos [ 2m(\xi (x) 
- \xi (y))] \chi_m(x,y)~.
\eeq
In the above expression, $V(x)$ is the bare potential of the 
impurities, $\xi$ is a phase factor which includes the 
$k_F x$ factor coming from the 
back-scattering process and other factors which arise due to the removal 
of the forward scattering terms from the Hamiltonian by shifts in the 
bosonic field $\phi$, and $\chi_m(x,y)$ is a factor which 
incorporates the renormalization group (RG) flows of the barrier strengths.
In general, it is given by a two-point correlation function defined as 
\beq
\chi_m(x,y) = \frac{1}{T^2} e^{-2m^2G_0(x,x,\tau_0) - 2m^2G_0(y,y,\tau_0)}
\int_0^{\infty} dt e^{4m^2G_0(x,y,it+\pi/2)}~,
\eeq 
$G_0 (x,y,it)$ is the two-point Green's function for a clean quantum wire. 
Here, the Green's functions are in terms of the imaginary time $\tau =it$. 
$\tau_0 \sim 1/\Lambda$ is the inverse of the high energy cutoff.
In a later subsection, we show how the one-point function 
$\chi_m(x,x)$ can be obtained directly from the RG equation for the barriers.

For our system,
we shall instead compute the two-point Green's function in terms of $\bar
\omega$, in terms of which, the correction to the conductance is given by 
(specializing to the case $m=1$) 
\bea 
R^{(1)} &=& {\rm lim}_{\omega \rightarrow 0} 
\frac{\omega}{(g_0 K_L)^2} \int dx' \int dy' G_0(x,x',\omega ) 
G_0(y,y',\omega)\cos [2(\xi (x') - \xi (y'))] \times \nonum \\
&& \quad \quad \quad \quad ~~~~~~~~~~~~~~~~~~~~ V(x') V(y') 
Im(F(x',y',\omega)-F(x',y',0)) \nonum \\
&=& \int dx' \int dy' \cos [2(\xi (x') - \xi (y'))] V(x') V(y')
{\rm lim}_{\omega \rightarrow 0} {dF(x',y',\omega) \over d \omega} ,
\eea
where 
\bea
&& \hspace*{-1cm} F(x',y',\omega) \nonum \\
&& \hspace*{-1cm}= \int_{-\infty}^{\infty} dt e^{i\omega t} \exp
( - 2\pi \sum_{\on'} [G_0(x',x',\on') + G_0(y',y', \on') - 2 G_0(x',y',\on')
\cos (\on' \tau)]).
\eea
Note that the Green's functions in the prefactors of $R^{(1)}$ are dependent
on the external driving frequency, but the Green's functions in
the exponential depend only on the Matsubara frequencies $\on'$, and not
on the external driving frequency $\omega$ or its analytic continuation
$\bo$. The sum over the Matsubara frequencies are cutoff at the low energy 
end by $ {\bar\omega}'_{n=1} \sim k_B T$ and at the upper end 
by the high energy cutoff $\Lambda$. In evaluating $R^{(1)}$, we will
approximate $\sum_{\on'}$ by $\int d\bo' /(2\pi)$ which is reasonable
since we always assume that the temperature $T$ is much smaller than the
cutoff $\Lambda$.

\subsection{Results for the Quantum Wire}

We will concentrate here on calculating the conductance of a quantum wire 
system in which the length of the quantum wire $l$ is much greater than 
the length of the contact regions $d$. Also, we will finally be interested 
in studying the effects of the junction barriers placed at the two 
lead-contact junctions (as explained earlier). Hence, for our model
\beq
V(x) = V_1 \Lambda \delta(x) + V_2 \Lambda \delta(x-L).
\eeq
For this potential, we can obtain the expression for the conductance 
corrections
as
\bea
R^{(1)} &=& \Lambda^{2} {\rm lim}_{\omega \rightarrow 0} 
[V_1^2 Im {dF(0,0,\omega)
\over d\omega} + V_2^2 Im {dF(L,L,\omega)\over d\omega} \nonum \\
&& \quad \quad ~~~~~+ 2V_1 V_2 Im {dF(0,L,\omega)\over d\omega}
\cos(\xi(0)-\xi(L))]. 
\label{r1}
\eea

\begin{itemize}

\item{} Spinless electrons

Now, the expression for the one-point Green's function (in frequency space) 
for a barrier placed inside 
the contact region on the left of the QW and at a distance $a$ from the left 
lead-contact junction (which is taken to be the origin, giving the hierarchy 
of length scales $a \ll d \ll l$) can be easily obtained from Appendix D.
It is given by
\beq
G_{\bar{\omega}} (x=a,y=a) \simeq \frac{K}{2\vert\bar{\omega}\vert} ~,
\eeq
where
\begin{displaymath}
K = \left\{ \begin{array}{ll}
K_C & \textrm{for $\vert\bar{\omega}\vert \gg v_C/a$}\\ \\
\frac{2K_LK_C}{K_L+K_C} & \textrm{for $v_C/d \ll \vert\bar{\omega}\vert 
\ll v_C/a$}\\ \\
\frac{2K_LK_W}{K_L+K_W} & \textrm{for $v_W/l \ll \vert\bar{\omega}\vert 
\ll v_C/d$}\\ \\
K_L & \textrm{for $\vert\bar{\omega}\vert \ll v_W/l$ ~.}
\end{array} \right.
\end{displaymath}
For our model with a barrier at the left lead-contact junction, $a=0$ and 
the first frequency regime does not exist.

The two-point Green's function $G(x,y)$ 
for $y$ in the left contact region and $x$ anywhere is given in 
Appendix D. By setting $y=0$ (i.e., at the first barrier) and $x=L=l+2d$
(i.e., at the second barrier) we 
obtain the conductances in the different frequency regimes given by
\bea
G &=& {2K_C(1+\gamma_1)^2 \over (2+K_C/K_W + K_W/K_C)} {\exp[-|\bo|
({2d\over v_C} +{l\over v_W})] \over |\bo|} \quad {\rm for}\quad 
|\bo| \gg v_W/l \gg v_C/d 
\nonum \\
&=& {2K_C(1+\gamma_1)^2 \over (2+K_C/K_W + K_W/K_C)} {\exp[-|\bo|
({2d\over v_C})] \over |\bo|} ~~\quad \quad \quad {\rm for} \quad 
v_W/l \gg |\bo| \gg v_C/d \nonum \\
&=& {K_L \over 2|\bo|} \quad \quad {\rm for} \quad v_C/d,v_W/l \gg |\bo| ,
\label{twopoint}
\eea
where $\gamma_1 = {K_L-K_C\over K_L+K_C}$. 
Thus, we see that $G(x,y, \bo')$ decays 
exponentially to zero except at the lowest frequency
regime where $G(x,y,\bo') = G(x,x,\bo') = G(y,y,\bo')$.

To obtain the conductance corrections, we
use the above Green's functions to compute $F(x',y',\bo')$ in each of these 
frequency regimes. For the high frequency regime, it is simply given by
\bea
F(x',y',\omega) &=& \int_{-\infty}^\infty dt 
e^{i\omega t} \exp [-2\pi \int_T^\Lambda {d\bo'\over 2\pi} 
({K_{eff} \over |\bo'|} - 2 G(x',y',\bo') \cos \bo' \tau )] \nonum \\
&=& \int_{-\infty}^\infty {1\over T} dz e^{i\omega z/T}
({T\over \Lambda})^{2K_{eff}} \exp [-2\pi \int_T^\Lambda
{d\bo'\over 2\pi} 2 G(x',y',\bo') \cos \bo' \tau] , \nonum \\
&&
\eea
where $K_{eff} = K_L K_C/(K_L + K_C)$, and 
in the second line, we have scaled $t$ by $T$, i.e., we
have used $t=z/T$ to write the integral in terms of dimensionless
variables so that the temperature power-laws can be made explicit.
When $x'\ne y'$, $G(x',y', \bo') \rightarrow 0$, so that one can check that
$lim_{\omega \rightarrow 0}Im {dF\over d\omega}$ also tends to zero.
This means that the cross-term in Eq. (\ref{r1}) does not contribute.
For each of the terms involving just one barrier, we find that 
\beq
lim_{\omega \rightarrow 0}Im {dF\over d\omega} = 
{1\over T^2}({T\over \Lambda})^{2K_{eff}} \exp [- \int_T^\Lambda
{d\bo'} {K_{eff}\over |\bo'|} \cos \bo' \tau].
\eeq 
Hence, we obtain the following answer for the conductance correction
for high temperatures $T\gg T_d \equiv v_C/(k_B d)$, 
\beq
g = g_0 K_L [ 1 - c_1 ({T\over \Lambda})^{2(K_{eff} - 1)}(|V_1|^2+ |V_2|^2)],
\label{hightemp}
\eeq
where $c_1$ is a dimensionful constant 
dependent on factors like the contact quasiparticle velocity $v_C$, but is
independent of the gate voltage $V_G$.
 
For intermediate temperatures 
where $T_l \equiv v_W/(k_B l) \ll T \ll T_d$ the calculation is very
similar to that performed for the high frequency case, except that the 
integral over the Matsubara frequencies is now split into two regions
\bea
F(x',y',\omega) &=& \int_{-\infty}^\infty dt 
e^{i\omega t} \exp [-2\pi (\int_T^{T_d} + \int_{T_d}^\Lambda)
{d\bo'\over 2\pi} {K \over |\bo'|} ]~ \nonum \\
&& \quad \quad \quad \exp
[2\pi \int_T^\Lambda {d\bo'\over 2\pi} 2 G(x',y', \bo') \cos \bo' \tau] ~.
\eea
The rest of the calculations go through as above, and we find that
the conductance expression is 
\beq
g = g_0 K_L [ 1 - c_2 ({T_d\over \Lambda})^{2(K_{eff} - 1)}
({T\over T_d})^{2({\tilde K}_{eff}-1)} (|V_1|^2 + |V_2|^2) ],
\eeq
where ${\tilde K}_{eff} = K_L K_W/(K_L + K_W)$ and $c_2$ is a dimensionful 
constant which is dependent on $V_G$. 

Finally, for very low temperatures $T \ll T_l$, the sum over 
Matsubara frequencies split into three regions so that we have
\bea
F(x',y',\omega) &=& \int_{-\infty}^\infty dt 
e^{i\omega t} \exp [-2\pi (\int_T^{T_l} + \int_{T_l}^{T_d}
+\int_{T_d}^\Lambda) {d\bo'\over 2\pi} {K \over |\bo'|} ] ~ \nonum \\
&& \quad \quad \quad \exp
[2\pi \int_T^\Lambda {d\bo'\over 2\pi} 2 G(x',y', \bo') \cos \bo' \tau] ~.
\eea
Furthermore, in this regime, the cross-term does not vanish; in
fact, for $x' \ne y'$, we have $G(x',y',\bo) = G(x',x',\bo)=G(y',y',\bo)$,
so that $F(x',y',\omega) =F(x',x',\omega) =F(y',y',\omega)$.
The contribution of the cross term is hence identical to that
of the terms due to a single barrier. Hence, we obtain the
corrections to the conductance as
\beq
g = g_0 K_L [ 1 - c_3 ({T\over T_l})^{2(K_L - 1)} ({T_d \over 
\Lambda})^{2(K_{eff} -1)} ({T_l\over T_d})^{2({\tilde K}_{eff} - 1)}
|V_1 + V_2|^2 ] ~,
\eeq
where $c_3$ is a dimensionful constant similar in nature to $c_2$; thus the 
two barriers are seen coherently.

Note that the power-laws come purely from the one-point Green's functions,
whereas the phase coherence between the barriers is determined by the
behavior of the two-point correlation function. At high or intermediate 
frequencies, $lim_{\omega \rightarrow 0} Im {dF (x',y',\omega) \over 
d\omega}$ tends to zero for $x'\ne y'$ leading to the lack of phase coherence 
between the two barriers. At very high temperatures, the 
interaction parameters of the contact region $K_C$ and the lead region $K_L$ 
controls the renormalization of a barrier in the contact region. 
As the temperature is lowered, the phase coherence length of the electronic 
excitations increases, and the renormalization exponent makes a crossover to 
a combination of the interaction parameters of the contact and QW, and finally 
to that of the lead alone at the lowest temperature regime. The lowest 
temperature regime is also the one in which resonant transport through both 
the lead-contact junction barriers can take place as phase coherence over the 
entire system is achieved at these temperatures.

\item{} Electrons with spin

The above expressions were given for a 
model of the QW system but for spinless electrons. Let us now see what the 
conductance expressions are for electrons with spin. These expressions can 
be derived in the same way as for spinless electrons by using the appropriate
Green's functions for spin and charge fields. This
gives us for the high temperature regime of $T_d \ll T $
\beq
g = 2 g_0 K_L [ 1 - c_4 ({T \over \Lambda})^{2(K_{eff} - 1)} 
(|V(0)|^2 + |V(l+2d)|^2) ],
\eeq
where now $K_{eff} = K_L K_{C\rho}/(K_L + K_{C\rho}) + K_L K_{C\sigma}/(K_L 
+ K_{C\sigma})$, and $c_4$ is a dimensionful constant much like $c_1$ for the 
spinless case (i.e., dependent on the contact charge velocity $v_{C,\rho}$ but 
independent of the gate voltage $V_G$). For the intermediate temperature range 
$T_l \ll T \ll T_d$, we find 
\beq
g = 2 g_0 K_L [ 1 - c_5 ({T_d \over \Lambda})^{2(K_{eff} - 1)}
({T \over T_d})^{2({\tilde K}_{eff}-1)} (|V(0)|^2 + |V(l+2d)|^2) ] ~,
\eeq
where ${\tilde K}_{eff} = K_L K_{W\rho}/(K_L + K_{W\rho}) + 
K_L K_{W\sigma}/(K_L + K_{W\sigma})$, and $c_5$ is a constant similar to 
$c_2$ for the spinless case (i.e., dependent on $V_G$). Finally, for the low 
temperature regime $T \ll T_l$, we obtain
\beq
g = 2 g_0 K_L [ 1 - c_6 ({T \over T_l})^{2(K_L - 1)} ({T_d \over \Lambda})^{2
(K_{eff} -1)} ({T_l \over T_d})^{2({\tilde K}_{eff} - 1)}
|V(0) + V(l+2d)|^2 ] ~,
\eeq
where $c_6$ is similar in nature to $c_3$ for the spinless case.

\end{itemize}

\subsection{Results for the Quantum Point Contact}

The Quantum Point Contact (QPC) is simply a quantum wire system in which the 
length of the quantum wire region $l \sim 0.2 - 0.5 \mu m$ (i.e., the region 
undergoing the constriction due to the application of the gate voltage) 
is much reduced in comparison to typical lengths for a quantum wire 
$l \sim 2 - 20 \mu m$. Thus, in our model of the quantum wire system, we can 
reach the QPC by studying the limit when the contact region length $d$ is much 
greater than the wire length $l$. Let us then study the effects of 
barriers/impurities placed in the contact and wire region of the QPC.

\begin{itemize}

\item{} Spinless electrons

In order to study the effect of a weak barrier placed in the contact 
region such that its distance $a$ from the left lead-contact junction falls 
in the hierarchy of $a \ll l \ll d$, we again start by computing the one-point 
Green's function for such an impurity. We find that
\beq
G_{\bar{\omega}} (x=a,y=a) \simeq \frac{K}{2\vert\bar{\omega}\vert},
\eeq
where
\begin{displaymath}
K = \left\{ \begin{array}{ll}
K_C & \textrm{for $\vert\bar{\omega}\vert \gg v_C/a$}\\ \\
\frac{2K_LK_C}{K_L+K_C} & \textrm{for $v_W/l \ll \vert\bar{\omega}\vert 
\ll v_C/a$}\\ \\
\frac{2K_LK_C}{K_L+K_C} & \textrm{for $v_C/d \ll \vert\bar{\omega}\vert 
\ll v_W/l$}\\ \\
K_L & \textrm{for $\vert\bar{\omega}\vert \ll v_C/d$ ~.}
\end{array} \right. 
\end{displaymath}

As before, for the two-point function, we find that the high and low
frequency limits are the same as that given in Eq. (\ref{twopoint}) for the 
QW, but for $T_d \ll T \ll T_l$, the answer turns out to be
the same as in the high frequency limit. This is similar
to what one sees for the one-point Green's functions above as well.
So, without giving any further derivations,
we directly quote the expressions for the conductance corrections.
In the high and intermediate frequency regimes, the conductance is given by
\beq
g = g_0 K_L [ 1 - c_i 
({T\over \Lambda})^{2(K_{eff} - 1)} (|V_1|^2+ |V_2|^2) ],
\eeq
where $i=4,5$ allows for the constant to be different in the high and 
intermediate frequency regimes. For the low frequency regime, we get
\beq
g = g_0 K_L [ 1 - c_6
({T\over T_d})^{2(K_L - 1)} ({T_d\over \Lambda})^{2(K_{eff} - 1)} 
(|V_1 + V_2|^2) ],
\eeq
It is clear from the above expressions 
that the contributions of barriers in the contacts of a QPC are always 
going to be independent of the gate voltage $V_G$ as the QPC interaction 
parameter $K_W$ does not enter anywhere. Thus, such an impurity would always 
lead to a flat and channel independent renormalized conductance. 
It should be noted that we have found from a similar calculation that even 
for an impurity placed deep inside the contact (i.e., with the hierarchy of 
$l \ll a \ll d$), the above conclusions still remain true; this is because 
the only change that takes place is that $K = K_C$ (rather than the 
combination of $K_L$ and $K_C$ found earlier) for the regime of $v_W/l \ll 
\vert\bar{\omega}\vert \ll v_C/a$.
 
Finally, let us study the effect of an impurity placed inside the QPC itself. 
We find the one-point Green's function for such a case (with the hierarchy of 
$l \ll d < a$) to be
\beq
G_{\bar{\omega}} (x=a,y=a) \simeq \frac{K}{2\vert\bar{\omega}\vert},
\eeq
where
\begin{displaymath}
K = \left\{ \begin{array}{ll}
K_W & \textrm{for $\vert\bar{\omega}\vert \gg v_W/l$}\\ \\
K_C & \textrm{for $v_C/d \ll \vert\bar{\omega}\vert 
\ll v_W/l$}\\ \\
K_L & \textrm{for $\vert\bar{\omega}\vert \ll v_C/a$ ~.}
\end{array} \right. 
\end{displaymath}
We have here only three frequency regimes as the regime of 
$v_C/d \ll \vert\bar{\omega}\vert \ll v_W/l$ cannot be taken sensibly within 
the given hierarchy of length scales. This shows again that the effect of an 
impurity placed within the QPC will always be dependent on the gate voltage 
$V_G$, and can never lead to flat and channel independent renormalizations of 
the conductance. 

\item{} Electrons with spin

The generalization to spinful electrons can be obtained just
as was done for quantum wires with the appropriate substitutions.

\end{itemize} 

\subsection {Evaluation of the conductances from the effective
actions using the RG equations}

Here, we note that the above results for the conductances
could have been anticipated by computing the RG equation for the impurity 
potentials using the effective actions calculated in Sec. 3.

The conductance is governed by the renormalized barrier potentials at the two 
junctions. Since the interaction is repulsive, the barrier potentials
are expected to grow as a function of the frequency cutoff. This is
what leads to the result that any impurity potential,
however small, eventually cuts the wire; in the zero temperature
limit, there is no transmission at all \cite{kane}. However, 
at a finite temperature $T$, finite wire length $l$ or finite contact
length $d$, the growth is cutoff by either $T$, $v_W/L$ or $v_C/d$.
In fact, since the energy scales in the problem are the 
temperature $k_BT$, the high frequency cutoff $\Lambda$ and those related 
to the contact length $k_BT_d= v_C/d$ and 
the wire length $k_BT_l= v_W/l$, we can see that there will 
exist two energy scale crossovers in the system --- one from $T/\Lambda$ 
to $T/T_d$ and the other from $T/T_d$ to $T/T_l$ for $d \ll l$ (the QW 
limit), or from $T/\Lambda$ to $T/T_l$ and then from $T/T_l$ to $T/T_d$ for 
$l \ll d$ (the QPC limit). 

In fact, an explicit RG calculation of either of the individual barrier
strengths in the high, intermediate and low frequency regimes 
simply involves computing the dimension
of $\cos (2\sqpi\phi_1)$ or $\cos (2\sqpi\phi_4)$ (which turn out to be 
the same) using those respective actions. For example, for the high 
frequency effective action given in Eq. (\ref{highfreq}), the RG 
equation for a single barrier is given by
\beq
{dV_1\over d\lambda} = (1 - {2K_C K_L\over K_C+K_L}) V_1 \equiv 
(1 - K_{eff})V_1 ~,
\eeq
where $\lambda={\rm ln} \frac{\Lambda(\lambda)}{\Lambda}$.
Using this, we can get the renormalized barrier strength to be
\beq
V_1^{ren} = V_1 ({T\over \Lambda})^{K_{eff}-1}
\eeq
in the high frequency regime $T\gg T_d$, where we have used $T$ to 
cutoff the RG flow (which begins from $\Lambda$). From this, we infer that
to quadratic order, the $T$ dependence of the conductance 
corrections is given by
\beq
g = {e^2\over h} [1 - {\tilde c}_1V_1^2 (\frac{T}{\Lambda})^{2(K_{eff}-1)}]
\quad {\rm for} \quad
T\gg T_d ~,
\label{offres}
\eeq 
where ${\tilde c}_1$ is a dimensionful constant like $c_1$ defined 
earlier containing factors like $v_C$ but 
is, most importantly, independent of the gate voltage $V_G$.
Comparing with Eq. (\ref{hightemp}), we see that if we include the 
subtraction due to two barriers, the expressions are identical.

In the intermediate regime of $T_l \ll T \ll T_d$, the RG equation for the 
same barrier now becomes
\beq
{dV_1\over d\lambda} = (1 - {2K_C K_W\over K_W+K_L}) V_1 \equiv
(1 - {\tilde K}_{eff})V_1 ~.
\eeq
using the effective action in Eq. (\ref{intertemp}).
At the same time, the appearance of the energy scale $v_C/d$ (through 
$U_C$) in the effective action in this temperature regime and the taking 
of the approximation $\on \ll U_C$ means that $v_C/d$ 
has replaced $\Lambda$ as the high energy cutoff in the expression for 
the $T$ dependence of the conductance correction. The influence of 
those degrees of freedom whose energies lie between $v_C/d$ and $\Lambda$ 
can be taken into account by noting that they will contribute a 
factor of $(T_d/\Lambda)^{2(K_{eff}-1)}$; this is because 
these degrees of freedom have been integrated away during the RG 
procedure, and there must be continuity between the 
conductance expressions for $T \gg T_d$ and $T_l \ll T \ll T_d$ at 
$T = T_d$. Thus, we get the conductance expression in this regime as
\beq
g = {e^2\over h} [1 - {\tilde c}_2V_1^2 (\frac{T_d}{\Lambda})^{2(K_{eff}-
1)}) (\frac{T}{T_d})^{2({\tilde K}_{eff}-1)}]
\quad {\rm for} \quad T_l\ll T\ll T_d ~,
\label{intcond}
\eeq
where ${\tilde c}_2$ is a constant similar in nature to ${\tilde c}_1$, but it 
can depend on $v_W$ and is hence dependent on the gate voltage $V_G$. Thus the 
conductance is no longer independent of $V_G$. This again is the same as the 
expression obtained by the explicit computation of the conductance.

Finally, in the low temperature limit, 
we recognize the fact that there is phase coherence over the distance 
between the two barriers; this follows from
the low frequency effective action which has cross terms between
the fields at the two barriers. This is what leads to resonant transmission. 
To compute the conductance corrections away from resonance in this limit, 
we note the following. Since the resonance occurs precisely when the 
$2k_F$ component of the barrier term goes to zero, the relevant
term away from this resonance is precisely the back-scattering potential
$V\cos (2\sqpi \chi)$. Computing the dimension of this operator gives us the 
RG equation for our barriers in this temperature regime as
\beq
{dV_1\over d\lambda} = (1 - K_L) V_1 .
\eeq
This makes the $T$ dependence of the conductance correction clear. Again, 
the appearance of the energy scale $v_W/l$ (through $U_W$) and 
the approximation $\on \ll U_W$ indicate 
that $v_W/l$ has now replaced $v_C/d$ as the high energy cutoff in the 
expression for the $T$ dependence of the conductance correction. 
As before, the influence of those degrees of freedom whose energies lie in
between $v_W/l$ and $v_C/d$ is shown by the appearance of the term 
$(T_l/T_d)^{2({\tilde K}_{eff}-1)}$. This is because these degrees 
of freedom have also been integrated away during the RG procedure, and 
there must be continuity in the conductance expressions at $T=T_l$ whether we 
come from the $T_d \gg T \gg T_l$ regime or the $T_l \ll T$ regime. Thus,
we obtain the conductance in this regime as
\beq
G= {e^2\over h} [1 - {\tilde c}_3V_1^2 (\frac{T_d}{\Lambda})^{2(K_{eff}-1)}) 
(\frac{T_l}{T_d})^{2({\tilde K}_{eff}-1)} (\frac{T}{T_l})^{2(K_L-1)}] 
\quad {\rm for} \quad T\gg T_L ~,
\eeq
where ${\tilde c}_3$ is a constant similar in nature to ${\tilde c}_2$. We can 
now see that, as $K_L=1$ (for 2DEG Fermi reservoirs), the conductance has no
temperature dependence in the low temperature regime. 

A similar analysis can be done for the QPC limit, which reproduces the 
conductance expressions for the QPC limit that were obtained explicitly in 
the earlier subsection. We note, however, that the conductance 
corrections are small in this case as the RG flow for the barriers 
is restricted by the small length scales in the system.

The conductance corrections for electrons with spin can also be
obtained using the effective actions and the RG equations for the 
barriers, by proceeding in the same way as was done for spinless 
electrons. Since the conductance expressions have already been
given in the previous section, we do not repeat them here.

Thus, we emphasize that just by using the effective action and the RG 
equations for the barriers, we can actually obtain the conductance 
corrections. However, all that we actually do here is
to compute the RG flows of the individual barriers, and then infer
the temperature and length power-laws in the
conductance corrections. Hence, even in principle, there is no way of 
obtaining the constants ${\tilde c}_1 ,\ldots ,{\tilde c}_6$ from this method,
whereas the explicit computation of the conductance in the earlier subsection
can give the explicit forms of the constants as well. 
In fact, the correlation functions computed there
can be directly related to the coefficients which appear in the RG
equations. In the various frequency regimes, the RG equation for a 
single barrier for spinless fermions can be written as 
\beq
{dV\over d\lambda} = (1-2 |\bo| G_\bo (x,x))V. 
\eeq 
On Fourier transforming, this gives
\beq
\frac{dV (x)}{d\lambda} = ( 1 + 2\frac{dG}{d\lambda}(x,x,\tau_0 e^\lambda))
V(x) ~,
\eeq
where $\tau_0$ is the high energy cutoff $1/\Lambda$. Integrating this 
gives the renormalized strength of the impurity $V_{ren}$ as
\beq
V_{ren}(x,\lambda) = V(x,\lambda=0) \exp [\lambda - U(x,x,\tau_0 
e^\lambda)] ~,
\eeq
where $U(x,x,\tau_0e^\lambda) = -2 (G_0(x,x,\tau_0 e^\lambda) - 
G_0(x,x,\tau_0))$.
Thus, in this case, 
\beq
\chi_1 (x,x) = \exp [l -U(x,x,\tau_0 e^\lambda )] ~.
\eeq
However, the non-local $\chi_1 (x,y)$ is not so easy to obtain just
from the RG equations.

Now, let us study the conclusions that can be drawn from the conductance
expressions. To begin with, the expressions in the various frequency regimes
reveal that as {\it either} the temperature $T$ 
is raised {\it or} the total length $L$ of the contacts and QW is decreased, 
the conductance corrections become smaller and the conductance approaches
integer multiples of $2g_0$ as expected \cite{tarucha,yacoby}. Furthermore, 
we can see from these expressions that in the high temperature limit
i.e., when $T \gg T_d, ~T_l$, the conductance corrections are independent 
of the QW parameters. Hence, they are {\it independent} of the gate voltage 
$V_G$ and of all factors dependent on the channel index. Thus they yield
renormalizations to the ideal values which are themselves plateau-like
and uniform for all channels. Such corrections to
the conductance explain some of the puzzling features observed in the
experiments of Ref. \cite{yacoby}. A more detailed 
comparison of these results against experimental findings will be made 
in a later section; it is important to note here that our results are in 
qualitative agreement with most experimental observations on electronic 
transport through a variety of quantum wire systems.

Let us now compare these observations with what we find as the 
perturbative renormalizations to the perfect conductance of a barrier/impurity 
placed anywhere within the quantum wire itself such that its distance from 
the left lead-contact junction (taken as the origin) is again denoted by $a$. 
An exactly similar computation of the one-point Green's function in this 
case reveals that (note that we are now working with the hierarchy of 
$d \ll a \ll l$)
\beq
G_{\bar{\omega}} (x=a,y=a) \simeq \frac{K}{2\vert\bar{\omega}\vert} ~,
\eeq
where
\begin{displaymath}
K = \left\{ \begin{array}{ll}
K_W & \textrm{for $\vert\bar{\omega}\vert \gg v_C/d$}\\ \\
K_W & \textrm{for $v_W/a \ll \vert\bar{\omega}\vert 
\ll v_C/d$}\\ \\
\frac{2K_LK_W}{K_L+K_W} & \textrm{for $v_W/l \ll \vert\bar{\omega}\vert 
\ll v_C/a$}\\ \\
K_L & \textrm{for $\vert\bar{\omega}\vert \ll v_W/l$ ~.} 
\end{array} \right.
\end{displaymath}
Now, for a barrier at the left contact-QW junction, $a \rightarrow d$,
 the second frequency regime of $v_W/a \ll \vert\bar{\omega}\vert 
\ll v_C/d$ does not exist. We also
find that $K = K'_{eff} \equiv 2K_CK_W/(K_C+K_W)$ rather than $K_W$ for
 $\vert\bar{\omega}\vert \gg v_C/d$.
Thus for a quantum wire system which has the two 
contact regions and only barriers at the two contact-QW junctions, the 
conductance in the highest temperature regime of $T \gg T_d$ is
\beq
g = g_0 K_L [ 1 - {\tilde c}_1 ({T \over \Lambda})^{2(K'_{eff} - 1)} 
(|V(d)|^2 + |V(l+d)|^2) ] ~.
\eeq
Here ${\tilde c}_1$ is a dimensionful 
constant which will depend on factors like $v_W$ and hence also the gate 
voltage $V_G$. For the intermediate temperature regime of $T_l \ll T \ll T_d$, 
we find the conductance to be
\beq 
g = g_0 K_L [ 1 - {\tilde c}_2 ({T_d \over \Lambda})^{2(K'_{eff} - 1)}
({T \over T_d})^{2({\tilde K}_{eff}-1)} (|V(d)|^2 + |V(l+d)|^2) ] ~,
\eeq
where ${\tilde K}_{eff} = 2K_L K_W/(K_L + K_W)$ as before, and ${\tilde c}_2$
is also a dimensionful constant dependent on $V_G$. Finally, for the lowest 
temperature regime of $T \ll T_l$, we obtain
\beq
g = g_0 K_L [ 1 - {\tilde c}_3 ({T \over T_l})^{2(K_L - 1)} ({T_d \over 
\Lambda})^{2(K'_{eff} - 1)} ({T_l \over T_d})^{2({\tilde K}_{eff} - 1)}
|V(0) + V(l+2d)|^2 ] ~,
\eeq
where ${\tilde c}_3$ too is a dimensionful constant dependent on $V_G$. Thus, 
we can see that any barrier or impurity placed anywhere inside the QW will 
always give a perturbative renormalization to the conductance which will be 
dependent on the gate voltage and hence can {\it never} be flat or even 
channel independent. The conductance expressions for the case of spinful 
electrons can be found for this case in exactly the same way as before.

\section{\bf Effects of a Magnetic Field}

In this section, we will study the effects of an in-plane magnetic field on 
the conductivity of a quantum wire. In general, a magnetic field couples to 
both the spin (Zeeman coupling) and the orbital motion of an electron. 
However, orbital motion is not possible in an in-plane magnetic field
because the electrons are constrained to move only in the plane.
Thus we will only consider the 
effect of the Zeeman term. This term couples differently to spin up 
and spin down electrons; here up and down are defined with respect to the 
direction of the magnetic field which may or not be parallel to the quantum 
wire. Thus the $SU(2)$ symmetry of rotations is explicitly broken. We will now 
see that the spin and charge degrees of freedom do not decouple any longer.
Our findings reveal that 

\noindent (a) for low magnetic fields (of about $0-3T$ for 
Ga-As systems), the Zeeman splitting of the Fermi energies of the 
two spin species of electrons in the QW is very small, and its effects 
can be ignored. 

\noindent (b) for intermediate magnetic fields (of about $3-8T$),
the Zeeman splitting of the Fermi energies of up and down spins becomes 
appreciable; up and down spins see the two barrier
strengths renormalize differently because of the Zeeman splitting, giving
rise to an odd-even effect in the conductance of the two spin species.

\noindent (c) at still higher magnetic fields (of about $8-16T$),
when each of the earlier sub-bands is completely Zeeman split into two
spin-split sub-bands, conductance steps will be seen in multiples of 
$g_0=e^2/h$; the odd-even effect will be most pronounced here with
odd numbered spin-split sub-bands (containing only aligned
moments) having a much less renormalized conductance and
even numbered spin-split sub-bands (containing only
anti-aligned moments) having a much more renormalized conductance, and 
we can treat each spin-split sub-band as an effectively spinless TLL system.

\noindent (d) at magnetic fields much higher than this, all the spins in the 
system will be spin polarized.
 
\subsection{The infinite TLL Quantum Wire and the Odd-Even Effect}

Let us first consider an infinitely long wire containing noninteracting 
electrons. A magnetic field $h$ contributes the following term to the 
Hamiltonian
\beq
- ~g \mu_B h ~S_{z,total} ~=~ - ~\frac{g \mu_B h}{2} ~(~ \rho_{0\ua} ~
+~ \frac{1}{\sqrt \pi} ~\partial_x \phi_\ua ~-~ \rho_{0\da} ~-~ 
\frac{1}{\sqrt \pi} ~\partial_x \phi_\da ~)~, 
\label{magham}
\eeq
where $g$ is the gyromagnetic ratio (which is $2$ for free electrons but
may be substantially smaller in quantum wire systems), $\rho_{0\ua} ,
\rho_{0\da}$ respectively denote the mean density of the spin up 
and spin down electrons, and $\phi_\ua , \phi_\da$ denote the 
bosonic fields for the spin up and down electrons. The density terms
$\rho_{0\sigma} = \rho_{0\ua} - \rho_{0\da}$ 
have a bigger effect than the derivative terms $\partial_x
\phi_\sigma$; by altering the chemical potentials for spin up and down
electrons, these terms lead to different Fermi momenta 
and therefore to different Fermi velocities $v_{F\ua}$
and $v_{F\ua}$ for the two kinds of electrons. 

We now add a density-density interaction of the form ${\cal U} \rho^2 /2$ 
where $\rho = \rho_\ua + \rho_\da$. (For instance, this may describe a 
short-range Coulomb repulsion as in the Hubbard model; in that case $\cal U$ 
is positive). The bosonized Lagrangian density takes the form
\bea
{\cal L} ~=~ & & \frac{1}{2v_{F\ua}} ~(\partial_t \phi_\ua )^2 ~-~
\frac{v_{F\ua}}{2} ~(\partial_x \phi_\ua )^2 ~+~ 
\frac{1}{2v_{F\da}} ~(\partial_t \phi_\da)^2 ~-~
\frac{v_{F\da}}{2} ~(\partial_t \phi_\da)^2 \nonum \\
& & -~ \frac{\cal U}{2\pi} ~(~\partial_x \phi_\ua ~+~ \partial_x 
\phi_\da ~)^2 ~,
\label{hamlag1}
\eea
where we have dropped some additive constants, and have only kept terms
which are quadratic in the fields. We can rediagonalize this Lagrangian by 
defining two new fields $\phi_i$, velocities $v_i$ and interaction parameters 
$K_i$ (where $i=+,-$), and a mixing angle $\gamma$, where 
\bea
\phi_+ &=& {\sqrt {K_+ v_+}} ~(\frac{1}{\sqrt {v_{F\ua}}} \cos 
\gamma ~\phi_\ua + \frac{1}{\sqrt {v_{F\da}}} \sin \gamma ~\phi_\da )
\equiv p \phi_\ua + q \phi_\da , \nonum \\
\phi_- &=& {\sqrt {K_- v_-}} ~( - \frac{1}{\sqrt {v_{F\ua}}} \sin 
\gamma ~\phi_\ua + \frac{1}{\sqrt {v_{F\da}}} \cos \gamma ~\phi_\da )
\equiv r \phi_\ua + s \phi_\da ,
\label{transf1}
\eea
and
\bea
v_+^2 ~+~ v_-^2 ~&=&~ v_{F\ua}^2 ~+~ v_{F\da}^2 ~+~ \frac{\cal U}{\pi} ~
(~ v_{F\ua} ~+~ v_{F\da} ~)~, \nonum \\
(~ v_+^2 ~-~ v_-^2 ~)~ \cos ~(2\gamma) ~&=&~ v_{F\ua}^2 ~-~ 
v_{F\da}^2 ~+~ \frac{\cal U}{\pi} ~ (~ v_{F\ua} ~-~ v_{F\da} ~)~, \nonum \\
(~ v_+^2 ~-~ v_-^2 ~)~ \sin ~(2\gamma) ~&=&~ \frac{2{\cal U}}{\pi} ~{\sqrt {
v_{F\ua} v_{F\da}}}~, \nonum \\
K_+ v_+ ~(~ \frac{\cos^2 \gamma}{v_{F\ua}} ~+~ \frac{\sin^2 
\gamma}{v_{F\da}} ~) ~&=& ~1 ~, \nonum \\
K_- v_- ~(~ \frac{\sin^2 \gamma}{v_{F\ua}} ~+~ \frac{\cos^2 
\gamma}{v_{F\da}} ~) ~&=& ~1 ~.
\label{transf2}
\eea
The Lagrangian density in Eq. (\ref{hamlag1}) then takes the decoupled form
\beq
{\cal L} ~=~ \frac{1}{2K_+ v_+} ~(\partial_t \phi_+ )^2 ~-~ \frac{v_+}{2K_+} ~
(\partial_x \phi_+ )^2 ~+~ \frac{1}{2K_- v_-} ~(\partial_t \phi_- )^2 ~-~
\frac{v_-}{2K_-} ~(\partial_t \phi_- )^2 ~.
\label{hamlag2}
\eeq
Thus the charge and spin degrees of freedom get mixed since the fields 
$\phi_+$ and $\phi_-$ which diagonalize the Lagrangian will generally be 
different from the fields $\phi_\rho = (\phi_\ua + \phi_\da)/\sqrt{2}$
and $\phi_\sigma = (\phi_\ua - \phi_\da)/\sqrt{2}$. Note that if the 
magnetic field $h$
is zero, then $v_{F\ua} = v_{F\da}$ and $\gamma = \pi /4$; $\phi_+$ and
$\phi_-$ are then identical (up to a sign) to the charge and spin fields 
$\phi_\rho$ and $\phi_\sigma$.

We will now present the RG equations for a weak $\delta$-function 
impurity placed at the origin. For this, we need to 
compute the scaling dimension of the impurity term 
in the Lagrangian. We first write the impurity term in terms of fermionic
fields $\psi_{\ua}(0)$ and $\psi_{\da}(0)$, and then in terms of 
the bosonic fields $\phi_{\ua}(0)$ and $\phi_{\da}(0)$ at the origin as
\bea
L_{imp} &=& V(0) (\psi_{\ua}^{\dag}(0) \psi_{\da}(0)
+ \psi_{\da}^{\dag}(0) \psi_{\ua}(0)) \nonum \\
&=& V_1 \Lambda ~[\cos(2\sqpi\phi_{\ua}(0)) +\cos(2\sqpi\phi_{\da}(0))] ~.
\eea
We then invert the relations between $\phi_{\pm}$ and 
$\phi_{\ua,\da}$ given above in order to rewrite the above 
expression for the impurity in terms of the diagonal fields $\phi_{\pm}$. 
The scaling dimensions of the impurity terms are then found to be
\beq
D_{V_{\ua}} = v_{F\ua} (\frac{\cos^{2}\gamma}{v_+} 
+ \frac{\sin^{2}\gamma}{v_-}) ~~~,~~~ D_{V_{\da}} = 
v_{F\da} (\frac{\sin^{2}\gamma}{v_+} + \frac{\cos^{2}\gamma}{v_-}).
\eeq
Hence the RG equations for $V_{\ua}$ and $V_{\da}$ are given by
\bea
\frac{dV_{\ua}}{d\lambda} &=& ( 1 - D_{V_{\ua}}) V_{\ua} \nonum \\
\frac{dV_{\da}}{d\lambda} &=& ( 1 - D_{V_{\da}}) V_{\da} ~,
\eea
where $V_\ua$ and $V_\da$ both start from the value $V_1$ at the microscopic
length scale. 

We can now study what happens in the presence of strong and 
weak magnetic fields. But let us first remind ourselves of the following 
relations (which result from the Zeeman splitting),
\bea
v_{F\ua,\da} &=& v_F \sqrt{1 \pm \frac{g\mu_{B} h}{2E_{F1D}}} \nonum \\
\tan(2\gamma) &=& \frac{2{\cal U} \sqrt{v_{F\ua}v_{F\da}}/\pi}
{(v_{F\ua}-v_{F\da})(v_{F\ua}+v_{F\da} + {\cal U}/\pi)} ~,
\eea 
where $v_F ={\sqrt {2E_{F1D}/m}}$ is the Fermi velocity in the absence of a 
magnetic field. Therefore, in the limit of a strong magnetic field where 
the Zeeman splitting of the two spin species is much larger 
than the short ranged interaction energy $\cal U$ (i.e., ${\cal U} \ll \vert 
v_{F\ua}-v_{F\da}\vert$ and $\gamma \ll \pi/2$), we can approximate 
the relations for the two velocities 
$v_{\pm}$ (to linear order in ${\cal U}/(v_{F\ua}-v_{F\da})$) as
\bea 
v_+ &=& v_{F\ua} + \delta_+ ~~~,~~~ \delta_+\ll v_{F\ua} \nonum \\
v_- &=& v_{F\da} + \delta_- ~~~,~~~ \delta_-\ll v_{F\da} ~.
\eea
By putting these relations for $v_{\pm}$ into the expressions given 
above relating $v_{\pm}$ and $v_{F\ua,\da}$, we get 
\beq 
\delta_+ ~=~ \delta_- ~=~ \frac{\cal U}{2\pi} ~.
\eeq
Now, using this (together with the fact that $\gamma \ll \frac{\pi}{2}$) in 
the two RG equations obtained above gives us
\bea
\frac{dV_{\ua}}{d\lambda} &=& (1 - \frac{v_{F\ua}}{v_+}) V_{\ua} \nonum \\
&\simeq& \frac{\cal U}{2\pi v_{F\ua}} V_{\ua} ~,
\eea
and
\bea
\frac{dV_{\da}}{d\lambda} &=& ( 1 - \frac{v_{F\da}}{v_-}) V_{\da} \nonum \\
&\simeq& \frac{\cal U}{2\pi v_{F\da}} V_{\da} ~.
\eea
Now, these two RG equations indicate that as $v_{F\ua}$ is larger than 
$v_{F\da}$, the renormalized impurity strength felt by those electrons 
which have their magnetic moments aligned with the external B field is less 
than the renormalized impurity strength felt by the electrons which have their 
magnetic moments anti-aligned with the B field. Furthermore, if the B field 
is further increased, we will reach a situation when alternate sub-bands 
in the QW will be populated by either only the aligned or only the 
anti-aligned electrons. In this regime, the difference in back-scattering 
felt by the two species of electrons will be very clear from the alternating 
weak and strong corrections to the conductance. To be more specific, all odd 
numbered sub-bands will show much less corrections to the perfect conductance 
(as they will be populated by electrons aligned with the magnetic field), 
while all even numbered 
sub-bands will show much greater corrections to the perfect conductance 
(as they will be populated by the anti-aligned electrons). This 
{\it odd-even} effect had, in fact, been predicted by a two-band TLL study of 
Kimura {\it et al} \cite{kimura}, but its explanation 
on the grounds of impurity 
renormalization is now made clear. Furthermore, this effect has been recently 
observed by Liang {\it et al} \cite{liang}, and we will discuss their 
observations in a later section. It should be noted here that though the 
odd-even effect is easy to show upon taking the limit of ${\cal U} \ll 
\vert v_{F\ua} 
- v_{F\da} \vert$, the existence of this phenomenon does not need 
this limit to be taken. Furthermore, we also find that upon taking the limit 
of $h \ll E_{F1D}$ (i.e., weak magnetic field with $v_{F\ua} = v_{F\da} =
v_F$), 
\beq
\frac{dV_{\ua}}{d\lambda} = \frac{dV_{\da}}{d\lambda} = \frac{\cal U}{2\pi 
v_F} ~,
\eeq
which tells us that the odd-even effect vanishes in the weak magnetic field 
limit. Finally, we comment on the fact that the odd-even effect discussed 
above gives rise to the possibility of the creation of a spin-valve (i.e., 
a spin polarized current creating device) in these QW systems. Even though 
the odd-even effect needs a high magnetic field to be observed in current 
day experiments \cite{liang}, it may be possible to employ artificial barriers 
like negatively-biased finger gates to heighten the difference in 
renormalization of the up and down spin electrons at lower magnetic fields. 
At this point, however, quantitative predictions are difficult to make.
 
\subsection{A study of our model for the QW with a magnetic field}

Having discussed how to obtain a diagonal Lagrangian when both a magnetic
field and interactions are present as well as shown the interesting odd-even 
effect that takes place because of an impurity in an infinite TLL in the 
presence of an external B field, we will now study what happens when the 
model in Sec. 2 is placed in a magnetic field. In the regions $x < 0$ and 
$x > l+2d$, we have a system of noninteracting electrons parametrized by 
velocities $v_{F\ua}$ and $v_{F\da}$. In the regions of the 
contacts $0<x<d$ and $l+d<x<l+2d$, we have an interacting system parametrized 
by two velocities $v_{C+}$, $v_{C-}$ and a mixing angle $\gamma_C$. In the
quantum wire $d<x<l+d$, the system is parametrized by the velocities $v_{W+}$, 
$v_{W-}$ and a mixing angle $\gamma_W$. The last six parameters are functions 
of $v_{F\ua}$, $v_{F\da}$ and the strengths of the interaction in the contacts
and quantum wire. The action for this model is given by
\bea
S_0 &=& \int d\tau ~[\int_{-\infty}^0 +\int_{l+2d}^\infty]~ dx ~\{ ~{1 \over 
2K_L}[ {1\over v_{F\ua}} (\pt\phi_\ua)^2 + v_{F\ua} (\px\phi_\ua)^2 ]~
+~ [\ua \rightarrow \da ] ~\} \nonum \\
&+& \int d\tau ~[\int_{0}^d +\int_{l+d}^{l+2d}] ~dx ~\{ ~{1 \over 2 K_{C+}}
[{1\over v_{C+}} (\pt\phi_{C+})^2 + v_{C+} (\px \phi_{C+})^2 ] ~+~ 
[ C+ \rightarrow C- ] ~\} \nonum \\
&+& \int d\tau ~\int_{d}^{l+d} ~dx ~\{ ~{1\over 2 K_{W+} }[{1\over v_{W+}} 
(\pt \phi_{W+})^2 + v_{W+} (\px\phi_{W+})^2 ] ~+~ [W+ \rightarrow W- ] ~\} ~. 
\nonum \\
&&
\label{actmag}
\eea
Note that we have 
ignored the junction barriers and the gate voltage for the moment. It is 
worth mentioning here that we have verified, by performing a calculation of 
the kind outlined in Ref. \cite{maslov}, that our model for the 
QW system when placed in an external magnetic field and in the absence of
any barriers/impurities still gives perfect conductance in the dc limit for 
each sub-band. 

We begin by noting that we will present here the calculation for the 
case when the mixing angle $\gamma$ is the same in both the contacts as well 
as the QW, i.e., the short ranged electron-electron interaction $\cal U$ is 
equal in all the three TLLs. Though this is not necessarily the case in a real 
system, we will present it as it considerably simplifies the computations 
while providing us with an adequate discussion of all the important results 
for effective actions, conductance expressions, resonances, etc. 
Later, we will briefly discuss the case in which
the mixing angle is different in the contact and QW regions. The explicit 
derivation of the effective action is given in Appendix C.
The high frequency effective Lagrangian density ($\bo \gg v_{C\pm}/d , 
v_{W\pm}/l$) simplifies to 
\bea
{\cal L}_{eff, high} &=& {|\bo| \over 2}[(\frac{1}{K_L} + 
\frac{p^2}{K_{C+}} + \frac{r^2}{K_{C-}})(\tphi_{1\ua}^2 + \tphi_{4\ua}^2) 
+ (\frac{1}{K_L} + \frac{q^2}{K_{C+}} + \frac{s^2}{K_{C-}})
(\tphi_{1\da}^2 + \tphi_{4\da}^2) \nonum \\
&&~~~~~ +~2(\frac{pq}{K_{C+}}+\frac{rs}{K_{C-}})(\tphi_{1\ua}
\tphi_{1\da} + \tphi_{4\ua}\tphi_{4\da})] \nonum \\
&& +~~{|\bo| \over 2}[\{p^2(\frac{1}{K_{C+}}+\frac{1}{K_{W+}}) +r^2(
\frac{1}{K_{C-}}+\frac{1}{K_{W-}})\}(\tphi_{2\ua}^2 +\tphi_{3\ua}^2) \nonum \\
&&~~~~~~~~~~ +~ \{q^2(\frac{1}{K_{C+}}+\frac{1}{K_{W+}}) +s^2(
\frac{1}{K_{C-}}+\frac{1}{K_{W-}})\}(\tphi_{2\da}^2 +\tphi_{3\da}^2) \nonum \\
&&~~~~~~~~~~ +~ 2\{pq(\frac{1}{K_{C+}}+\frac{1}{K_{W+}})
+rs(\frac{1}{K_{C-}}+\frac{1}{K_{W-}})\}(\tphi_{2\ua}\tphi_{2\da} 
+ \tphi_{3\ua}\tphi_{3\da})] \nonum \\
&& +~~ {\cal L}_{gate}(\phi_{2\ua,\da},
\phi_{3\ua,\da}) + {\cal L}_{imp}(\phi_{1\ua,\da}, \phi_{4\ua,\da}) .
\eea 
We can clearly see the separation between the outer two and inner two fields. 
This means that we can integrate out the inner fields 
$\tphi_{2\ua,\da}$ and $\tphi_{3\ua,\da}$ without any 
further work and be left with a high frequency effective action dependent 
on $\tphi_{1\ua,\da}$ and $\tphi_{4\ua,\da}$ exactly as 
given above (and without any influence of the gate voltage $V_G$ either). 
We can also make the prediction that the conductance corrections due to 
barriers at the outer two junctions will have temperature power-laws which 
will be combinations of $K_L$ and $K_{C\pm}$ (much like those seen before) and 
also that it will be not be dependent on the gate voltage. 

In the intermediate frequency range of $v_{W\pm}/l \ll \bo \ll v_{C\pm}/d$, we 
get (after integrating out the inner fields $\tphi_{2\ua,\da}$ and 
$\tphi_{3\ua,\da}$), 
\bea
{\cal L}_{eff,int} &=& {|\bo| \over 2}[(\frac{1}{K_L} + \frac{p^2}{K_{W+}} 
+ \frac{r^2}{K_{W-}})(\tphi_{1\ua}^2 + \tphi_{4\ua}^2) 
+ (\frac{1}{K_L} + \frac{q^2}{K_{W+}} + \frac{s^2}{K_{W-}})(\tphi_{1\da}^2 
+ \tphi_{4\da}^2) \nonum \\
&&~~~~ +~2(\frac{pq}{K_{W+}}+\frac{rs}{K_{W-}})(\tphi_{1\ua}
\tphi_{1\da} + \tphi_{4\ua}\tphi_{4\da})] \nonum \\
&& + \frac{{\tilde V}_G (p-q)(s-r)}{(ps-qr)}(\phi_{4\ua} - \phi_{1\ua} +
\phi_{4\da} - \phi_{1\da}) ~,
\eea
where $\alpha_{C\pm} = v_{C\pm}/(K_{C\pm}d)$ are the charging energies for 
the $\tphi_{\pm}$ fields, and they appear because of the growth of the 
coherence in the system over the contact regions. We can now predict that 
the conductance corrections due to barriers at the outer two junctions will
have temperature power-laws which will be combinations of $K_L$ and 
$K_{W\pm}$ (again like those seen previously), and that this correction will 
definitely be dependent on the gate voltage. 

Finally, in the low frequency 
limit of $\bo \ll v_{W\pm}/l$, we get
\bea
{\cal L}_{eff,low} &=& {|\bo| \over 2K_L} (\tphi_{1\ua}^2 + \tphi_{1\da}^2 + 
\tphi_{4\ua}^2 +\tphi_{4\da}^2) \nonum \\ 
&& +\sum_{\pm} {v_{C\pm}\over 2d}[\phi_{2\pm}^2 +\phi_{3\pm}^2 +
\phi_{1\pm}^2 + \phi_{4\pm}^2 -2\phi_{2\pm} \phi_{1\pm} -2\phi_{3\pm} 
\phi_{4\pm}] \nonum \\
&& +\sum_\pm {v_{W\pm}^2\over 2l v_{C\pm}}(\phi_{2\pm} -\phi_{3\pm})^2 
\nonum \\
&& +\sum_\pm \frac{{\tilde V}_G}{2(ps-qr)}[(s-r)(\phi_{3+}-\phi_{2+}) + 
(p-q)((\phi_{3-}-\phi_{2-})] .
\eea
Integrating out the inner fields, we are then left with an effective action in 
terms of the new current and charge variables respectively 
\bea
\phi_{14\ua,\da} &=& \frac{\phi_{1\ua,\da} + \phi_{4\ua,\da}}{2} \nonum \\
n_{14\ua,\da} &=& \frac{\phi_{4\ua,\da} - \phi_{1\ua,\da}}{\sqpi} ~.
\eea
that span the coherent TLL system
between the two lead-contact junction barriers in this 
low frequency regime. Thus, we get the effective Lagrangian density in this 
regime as
\bea
{\cal L}_{eff,low} &=& {|\bo| \over K_L}[\tphi_{14\ua}^2 + \tphi_{14\da}^2 + 
\frac{\pi}{4} {\tilde n}_{14\ua}^2 + \frac{\pi}{4} {\tilde n}_{14\da}^2] 
\nonum \\
&& + \frac{U_1}{2}(p n_{14\ua} + q n_{14\da} + n_{01})^2 + \frac{U_2}{2}(r 
n_{14\ua} + s n_{14\da} + n_{02})^2 + {\cal L}_{imp},
\eea
where 
\bea U_1 &=& \frac{\pi v_{W+}\alpha_{+}\beta_{+}} {v_{C+}\alpha_+ 
+ 2v_{W+}\beta_+} ~~,~~ U_2 = \frac{\pi v_{W-}\alpha_{-}\beta_{-}}
{v_{C-}\alpha_- + 2v_{W-}\beta_-} \nonum \\
n_{01} &=& \frac{{\tilde V}_G v_{C+}(s-r)}{\pi v_{W+}\beta_+ (ps-qr)} ~~,~~ 
n_{02} = \frac{{\tilde V}_G v_{C-}(p-q)}{\pi v_{W-}\beta_- (ps-qr)} \nonum \\
{\rm and} \quad \alpha_{\pm} &=& \frac{v_{C\pm}}{K_{C\pm}d} ~~,~~ 
\beta_{\pm} = \frac{v_{W\pm}}{K_{W\pm}d},
\eea
and the term coming from the two barriers can be written as
\beq
{\cal L}_{imp} = 2V \Bigl( \cos(2\sqpi\phi_{14\ua}) \cos(\pi n_{14\ua}) + 
\cos(2 \sqpi\phi_{14\da}) \cos(\pi n_{14\da}) \Bigr) ~.
\eeq
It becomes clear from the above expression that the temperature power-law 
will be dependent on the lead interaction parameter $K_L = 1$, and the 
conductance correction will, in this regime, be temperature independent (as 
seen previously). Furthermore, the length corrections will be gate 
voltage dependent.
Let us now study the possible resonance symmetries of the low frequency 
effective action given above. Even though the structure of this expression 
is more complicated than those encountered previously, we can rewrite the 
two charging terms as follows:
\beq 
{\cal L}_{charging} = \frac{(U_1 p^2 + U_2 r^2)}{2} [n_{14\ua} 
+ a n_{14\da} -b]^2 + f(n_{14\da}) + {\rm const} ~,
\eeq
where 
\bea
a &=& \frac{U_1 pq + U_2 rs}{U_1 p^2 + U_2 r^2} \nonum \\
b &=& \frac{U_1 n_{01} + U_2 n_{02}}{U_1 p^2 + U_2 r^2} ~,
\eea
and $f(n_{14\da})$ is a quadratic function of the field 
$n_{14\da}$ only. Thus, we can see that if we set 
\beq
a n_{14\da} - b = - Z - \frac{1}{2} ~,
\eeq
we get a resonance in the $n_{14\ua}$ parts of the charging and barrier 
terms whenever $n_{14\ua} = Z$ or $Z+1$ and one makes the transformation 
of $\phi_{14\ua}\rightarrow \phi_{14\ua}\pm \sqpi/2$. This means that 
only the transport of all up-spin electrons through the two barriers 
is at resonance, and this is clearly a one-parameter tuned resonance. 
A one-parameter tuned resonance for only the transport of all down-spin 
electrons through the two barriers can be found in exactly the same manner 
by rewriting the above charging expressions but for $n_{14\da}$ 
instead of $n_{14\ua}$. We also find another resonance given by
\bea
\frac{q}{p}n_{14\da} + \frac{1}{p}n_{01} &=& - Z - \frac{1}{2} ~, \nonum \\ 
{\rm and} \quad \frac{s}{r}n_{14\da} + \frac{1}{r}n_{02} &=& - Z - 
\frac{1}{2} ~,
\eea 
where $Z$ is the same integer in both equations; then there exists a possible 
resonance whenever $n_{14\ua} = Z$ or $Z+1$, and one makes the 
transformation of $\phi_{14\ua,\da}\rightarrow
\phi_{14\ua,\da}\pm \sqpi/2$. This resonance symmetry would 
lead to a vanishing of the barrier terms 
in the effective action, and would 
correspond to the transfer of an electron across the system. One 
can immediately see that the above two conditions on $n_{14\da}$ 
mean that two parameters, here the gate voltage $V_G$ and the external 
magnetic field $h$, have to be manipulated to achieve this resonance 
condition. Such a resonance will, therefore, be much more difficult to 
observe experimentally. However, this resonance will lead to a complete 
vanishing of all back-scattering (and hence the conductance corrections) 
while the two one-parameter tuned resonances will give only partial 
lessening of the conductance corrections. 

The two one-parameter tuned resonances can prove useful in 
creating a spin-valve. Even at a low magnetic field, the odd-even effect 
can be enhanced by using stronger artificial barriers (e.g., by
employing finger gates over the channel) or making the length of the channel 
longer or working at lower temperatures, together with tuning the
transport of only one spin species of electrons through the two barriers at 
resonance. Thus one can create an enhanced spin polarized electron current 
output from the QW system.

Before we go on to computing the conductance through the system for the above 
model in the presence of the magnetic field, let us make some remarks about 
the case when the mixing angle in the QW is taken to be different from that 
in the contacts. A long calculation does give expressions for the 
effective action in the three frequency regimes similar to those obtained 
above, but with two sets of the transformation coefficients 
relating the $\phi_{\pm}$ and $\phi_{\ua,\da}$ fields. However, 
the integrating out of the inner fields is a far more difficult task; 
furthermore, the analysis reveals that the only possible resonance symmetry 
of the low frequency effective action is one that needs at least four 
parameters to be manipulated. We will, therefore, not present these 
results as we do not find anything substantially 
new from the analysis compared to the simpler case of equal mixing angles.

\subsection{Conductance of our model for the QW with a magnetic field}

We will begin by a re-writing the RG equation, obtained by Safi and Schulz 
\cite{safilong} for an impurity placed within a QW of a finite size and 
connected to Fermi leads, in a way which will be convenient to use 
in computing the conductance expressions for our model of the QW with 
barriers even in the presence of the external magnetic field. We begin by 
quoting the expression for the RG flow found for an impurity in Ref.
\cite{safilong},
\beq
\frac{dV_{m_{\rho},m_{\sigma}}(x)}{d\lambda} = [1 - \frac{1}{2}(m_{\rho}^2 
\frac{dU_{\rho} (x,x,\tau_0 e^\lambda)}{d\lambda} + m_{\sigma}^2 \frac{dU_{
\sigma} (x,x,\tau_0 e^\lambda)}{d\lambda})] V_{m_{\rho},m_{\sigma}}(x),
\eeq
where
\bea
U_{\rho}(x,x,\tau_0 e^\lambda) &=& 2 [G_{\rho}(x,x,\tau_0) - G_{\rho}
(x,x,\tau_0 e^\lambda)] \nonum \\
U_{\sigma}(x,x,\tau_0 e^\lambda) &=& 2 [G_{\sigma}(x,x,\tau_0) 
- G_{\sigma}(x,x,\tau_0 e^\lambda)].
\eea
Now with $\phi_{\rho} = (\phi_{\ua} + \phi_{\da})/{\sqrt 2}$ and 
$\phi_{\sigma} = (\phi_{\ua} - \phi_{\da})/{\sqrt 2}$, we can write
\bea
G_{\rho}(x,x,\tau_0 e^\lambda) &=& <\phi_{\rho}(x,\tau_0 e^\lambda)
\phi_{\rho}(x,0)> = \frac{1}{2}[<\phi_{\ua}\phi_{\ua}> + <\phi_{\da}
\phi_{\da}> + 2<\phi_{\ua}\phi_{\da}>] \nonum \\
G_{\sigma}(x,x,\tau_0 e^\lambda) &=& <\phi_{\sigma}(x,\tau_0 e^\lambda)
\phi_{\sigma}(x,0)> = \frac{1}{2}[<\phi_{\ua}\phi_{\ua}> + <\phi_{\da}
\phi_{\da}> - 2<\phi_{\ua}\phi_{\da}>], \nonum \\
&&
\eea
where the space-time indices are implicit on the right hand sides. 
Substituting the expressions 
for $U_{\rho}$ and $U_{\sigma}$ given above in the RG equation, and 
working with the case for the back-scattering of one electron 
$m_{\rho} = m_{\sigma} = 1$, we write the RG equation as
\beq
\frac{dV_{1,1}(x)}{d\lambda} = [1 + \frac{d}{d\lambda}(G_{\ua}(x,x,\tau_0 
e^\lambda) + G_{\da}(x,x,\tau_0 e^\lambda))] V_{1,1}(x) ~,
\eeq
where $G_{\ua} = <\phi_{\ua}\phi_{\ua}>$ and 
$G_{\da} = <\phi_{\da}\phi_{\da}>$. We will now
use the effective actions found in the various frequency regimes 
to obtain the two Green's functions $G_{\ua}$ and $G_{\da}$, 
put these in the RG equations and thereby infer the corrections 
to the conductance caused by the junction barriers. 

We start with the high and intermediate frequency/temperature effective 
actions given earlier for the model when the mixing angle is the same in the 
contact and QW regions. Here, we can 
see that the final effective action (in terms of only the fields at the outer 
two junctions) is the sum of two distinct parts, each of which is an 
expression of the kind $\frac{A}{2}\phi_{\ua}^2 + 
\frac{B}{2}\phi_{\da}^2 + C\phi_{\ua}\phi_{\da}$ 
separately for fields $\phi_1$ and $\phi_4$. This tells us that we can simply 
take the sum of the contributions from each of the two incoherent barriers. 
Thus, the general expression 
\beq
{\cal L}_{eff} = \frac{A}{2\vert\bo\vert}\tphi_{\ua}^2 + \frac{B}{2\vert\bo
\vert}\tphi_{\da}^2 + \frac{C}{\vert\bo\vert}\tphi_{\ua}\tphi_{\da},
\eeq
can be diagonalized in terms of two new fields $\phi_a$ and 
$\phi_b$ i.e., written as
\beq
{\cal L}_{eff} = \frac{1}{2\vert\bo\vert} (\lambda_a {\tphi_a}^2 
+ \lambda_b {\tphi_b}^2) ~,
\eeq
where $\lambda_a$ and $\lambda_b$ are the eigenvalues of the transformation
given by
\beq
\lambda_{a,b} = \frac{A+B}{2} \pm \frac{1}{2}\sqrt{(A-B)^2+C^2} ~.
\eeq 
Then, 
\bea
G_{\bo a} &=& <\tphi_a \tphi_a> = \frac{1}{\lambda_a 
\vert\bo\vert} \nonum \\
G_{\bo b} &=& <\tphi_b \tphi_b> = \frac{1}{\lambda_b \vert\bo\vert}.
\eea
Using the two eigenvectors corresponding to these two eigenvalues, we obtain 
$G_{\bo\ua}$ and $G_{\bo\da}$ as
\beq
G_{\bo\ua} = \frac{C^2}{\vert\bo\vert \{(\lambda_a-A)(\lambda_b-B)-C^2\}^2} 
[C^2 G_{\bo a} + (\lambda_b - B)^2 G_{\bo b}] ,
\eeq
and
\beq
G_{\bo\da} = \frac{C^2}{\vert\bo\vert \{(\lambda_a-A) (\lambda_b-B)-C^2\}^2} 
[(\lambda_a-A)^2 G_{\bo a} + C^2 G_{\bo b}].
\eeq
We finally obtain an expression for $G_{\bo\ua} + G_{\bo\da}$ as
\bea
G_{\bo\ua} + G_{\bo\da} &=& \frac{C^2}{\vert\bo\vert \{(\lambda_a-A)
(\lambda_b-B)-C^2\}^2} [\frac{C^2 + (\lambda_a - A)^2}{\lambda_a}
+ \frac{C^2 + (\lambda_b - B)^2}{\lambda_b}] \nonum \\
&\equiv& \frac{K_{mag}}{2\vert\bo\vert} .
\eea
We can now use the Fourier transform of the above expression to obtain 
the temperature and length power-laws for the conductance corrections in 
the high and 
intermediate frequency regimes. In the high temperature regime of 
$T \gg T_d ~(\sim v_{C\pm}/d)$, we get the conductance as 
\beq
g = g_0 K_L [1 - c_1(|V(0)|^2 + |V(l+2d)|^2)(\frac{T}{\Lambda})^{(K_{eff,mag} 
- 2)}] ,
\eeq
where $c_1$ is a dimensionful constant independent of the gate voltage $V_G$,
and $K_{eff,mag}$ is given by the expression for $K_{mag}$ where 
the coefficients $A$, $B$ and $C$ are given by
\bea
A &=& \frac{1}{K_L} + \frac{p^2}{K_{C+}} + \frac{r^2}{K_{C-}} \nonum \\
B &=& \frac{1}{K_L} + \frac{q^2}{K_{C+}} + \frac{s^2}{K_{C-}} \nonum \\
C &=& \frac{pq}{K_{C+}} + \frac{rs}{K_{C-}} ~.
\eea
Similarly, the conductance expression for the intermediate temperature defined 
by $T_l ~(\sim v_{W\pm}/l) \ll T \ll T_d$ is given by 
\beq
g = g_0 K_L [1 - c_2(|V(0)|^2 + |V(l+2d)|^2)(\frac{{\tilde T}_d}{\Lambda})^{
(K_{eff,mag} -2)}(\frac{T}{{\tilde T}_d}) ^{({\tilde K}_{eff,mag} - 2)}],
\eeq
where $c_2$ is another dimensionful constant which is dependent on the gate 
voltage, $T_d$ has replaced $\Lambda$ as the correct cutoff for the 
temperature, and ${\tilde K}_{eff,mag}$ is found in exactly the same way as 
$K_{eff,mag}$ but with coefficients $A_1$, $B_1$ and $C_1$ defined as
\bea
A_1 &=& \frac{1}{K_L} + \frac{p^2}{K_{W+}} + \frac{r^2}{K_{W-}} \nonum \\
B_1 &=& \frac{1}{K_L} + \frac{q^2}{K_{W+}} + \frac{s^2}{K_{W-}} \nonum \\
C_1 &=& \frac{pq}{K_{W+}} + \frac{rs}{K_{W-}}.
\eea
Finally, we obtain the low frequency conductance expression for the 
temperature regime of $T \ll T_l$ as 
\beq
g = g_0 K_L [1 - c_3(|V(0) + V(l+2d)|^2)(\frac{{\tilde T}_d}{\Lambda})^{
(K_{eff,mag} -2)}(\frac{T_l}{T_d})^{({\tilde K}_{eff,mag} - 2)}
(\frac{T}{T_l})^{2(K_L - 1)}],
\eeq
where $c_3$ is a dimensionful constant similar in nature to $c_2$, i.e., 
dependent on gate voltage. This expression is also independent of the 
temperature for Fermi leads with $K_L = 1$, and the coherence between the 
barriers means that this correction term could go to zero at resonance. 

We end by noting that we can again take the limit of ${\cal U} \ll 
\vert v_{C\ua} - v_{C\da} \vert$ in our equations to highlight the 
existence of the odd-even effect within our model of the QW as well. Upon 
taking this limit in the high temperature regime, we find that
\beq
\frac{dV_{\ua}}{d\lambda} \simeq \frac{\cal U}{4\pi v_{C\ua}},
\eeq
while
\beq
\frac{dV_{\da}}{d\lambda} \simeq \frac{\cal U}{4\pi v_{C\da}},
\eeq 
where $\cal U$ is the inter-electron interaction term. This clearly shows 
that as $v_{C\ua}$ increases and $v_{C\da}$ 
decreases with an increasing magnetic field, the renormalized barrier seen 
by the two spin species of electrons will be different. We also note that, 
just like the case of the infinite, homogeneous QW, a weak field of $h \ll 
E_{F1D}$ in our model of the QW does not give rise to the odd-even effect. 

In summary, we can see that by turning on an external magnetic field in the 
QW system, the up and down spin electrons see different renormalized strengths 
of any barriers (or impurities) --- this is the odd-even effect. We speculate 
on the possible use of this effect in creating a spin-valve using QW 
systems. The effective actions, their resonance symmetries as well as the 
temperature and length power-law corrections to the conductance in the 
various temperature regimes, however, still follow a pattern similar to 
that for a QW without a magnetic field. 
 
\section{\bf Comparison with the Experiments}

We now discuss the relevance of this model to many of the experiments that 
have been performed so far on quantum wire systems fabricated using 
cleaved-edge overgrowth as well as split-gate techniques. But before 
doing that, let us reiterate some well-known
observations about the experimental system that we are trying to 
model here. In this system, the electrons enter the wire from the 
2DEG reservoirs lying outside the wire with 
a Fermi energy $E_F$ whose value (typically around $5 - 10meV$) is fixed by 
the parameters of the 2DEG. 
Within the quantum wire, the gate voltage produces a discrete set of sub-bands 
labeled by an integer $s$ (see Fig. 2); let $E_s$ denote the energies of the 
bottoms of these sub-bands. In a sufficiently long quantum wire, we expect 
$E_s$ to be constant along the length of the wire provided we are not too 
close to either of the junctions. Thus an electron which has energy $E_F$ 
and enters the sub-band $s$ will have a wave number $k_{Fs}$ inside the 
wire given by $k_{Fs}^2 /2m = E_F - E_s$ and a velocity given by 
$v_{Fs} = k_{Fs} /m$ \cite{buttiker}.
We know that if $N$ of the $1D$ sub-bands lie below the 2DEG
Fermi energy $E_{F2D}$ (which itself at any finite temperature is 
surrounded by a small thermal spread), we will get 
$N$ quantized steps in the conductance when the quantum wire is completely 
free of any impurities; this statement is true irrespective of the electron 
velocities, densities or how they interact among themselves while in the 
various channels \cite{safi,maslov,buttiker}. 
Now, upon increasing the gate voltage $V_G$, one adds an 
energy $eV_G$ to every electron in each of the $1$D sub-bands in the quantum 
wire. This has the effect of pushing up each of the sub-bands 
by the same energy and can even de-populate the 
sub-bands by pushing them above $E_{F2D}$ (see Fig. 2). Thus, changing 
the gate voltage decreases the electron density in the quantum wire and 
allows the transport process to take place through only a few channels, and 
in the extreme limit, only one channel, before cutting off the wire altogether 
by pushing all the $1$D sub-bands above the $E_{F2D}$ (this is called 
{\it pinch-off}). The conductance measurement which 
shows step quantization in terms of rises and plateaus can then be 
explained in the following way. Whenever, by decreasing the 
gate voltage $V_G$, the bottom one of the 1D sub-bands (which is initially 
well above $E_{F2D}$) first touches the top of the thermal spread just 
above $E_{F2D}$, that band starts filling up and so 
we can see a rise. Once the bottom of this sub-band 
crosses the bottom of the thermal spread just below $E_{F2D}$, the rise 
is topped off by a plateau which signals that another channel is fully open 
to electron transport between the two reservoirs (see Fig. 2). Some of the 
earliest experiments with quantum wires free of impurities did indeed reveal
quantization of the conductance in integer steps of $2g_0$ \cite{wees}.

But later Tarucha {\it et al} \cite{tarucha} performed experiments 
with wires of lengths of $2\mu m$ to $10\mu m$ fabricated using split-gate 
methods at temperatures from $0.3K$ to $1.1K$,
and found deviations from the perfect quantization of the steps. Attempts
were then made to explain these deviations as due to electron-electron
interactions. Although, a clean TLL wire between Fermi liquid
leads would not lead to renormalization of the conductance
quantization, several authors \cite{furusaki,safilong,maslovlong}, showed 
that the presence of impurities in a 
TLL connected to Fermi leads would cause renormalization. 
However, they expected the renormalizations to be gate voltage
dependent; this was indeed seen by Tarucha {\it et al} \cite{tarucha}.

However, Yacoby {\it et al} \cite{yacoby} made the following surprising 
observation for a quantum wire $2\mu m$ long fabricated in 
cleaved-edge overgrowth systems: the dc conductance 
showed several nearly flat plateaus whose 
heights are uniformly renormalized from the ideal 
values of integer multiples of $2g_0$ for measurements made over a 
temperature range of $0.3 - 25K$. Similar observations were subsequently 
made in several experiments on quantum wires made using the split-gate 
technique \cite{facer,liang,liang2,reilly}. In all these experiments
the step heights were increasingly renormalized as either the temperature 
was lowered (for a fixed length of quantum wire) or the length of the 
quantum wire was increased at a fixed temperature.
Such renormalizations would require back-scattering of electrons.
If these back-scatterings were due to impurities within
the quantum wires, the conductance corrections would be gate
voltage dependent as shown in our calculations. This can
certainly not lead to flat conductance plateaus as seen in the experiments. 

Our model, however, has contact regions independent of the gate voltage and 
has barriers at the contacts arising due to the changes in the
nature of the electron-electron interactions and geometry. 
Thus, the back-scattering at these barriers is independent of gate
voltage and the sub-band index (as can be seen in our results), and will 
lead to conductance plateaus which are flat as a function of 
gate voltage and uniform for all the sub-bands at the highest temperatures. 
We note that a recent experiment 
\cite{picciotto} on a quantum wire system similar to that used by Yacoby 
{\it et al} \cite{yacoby} revealed the existence of a 
region of length $2 - 6\mu m$ which lies in between the gated quantum wire 
region and the 2DEG reservoirs and gives rise to the back-scattering 
that causes the flat and uniform renormalization of the conductance of each 
sub-band. Such contact regions
correspond to $T_d \sim 0.2 - 0.7 K$. This is much less than most 
of the temperature range shown in Fig. 3 of Ref. \cite{yacoby}. 
The similar flat and uniform conductance corrections
seen in the experiments of Refs. \cite{facer,liang,liang2,reilly} 
seem to suggest that their QW systems also include contact regions and have
$T \gg T_d$. 

Now, as explained earlier for a quantum wire 
system in which the contact length $d \ll l$, in the intermediate and 
low temperature regimes of $T_l \ll T \ll T_d$ and $T \ll T_l$, 
we know that the correct cutoffs for the RG procedure are 
$T_d$ and $T_l$ respectively; that is why the length power-laws of 
$d$ and $l$ appear in the conductance corrections in these two regimes 
besides the customary temperature power-law. We can clearly see that the 
inverse length scale $d^{-1}$ (for the contact region) and $l^{-1}$ (for 
the wire region) have similar power-laws to those obtained for the 
temperature. Thus, one 
can qualitatively understand the increase in the conductance with increasing 
temperature and its decrease as the length of the quantum wire is increased.
This has been observed by several groups \cite{tarucha,yacoby,facer,liang2}. 
Furthermore, one recent experiment using a split-gate QW system 
\cite{reilly} shows that the conductance of a $2\mu m$ long QW
at $T = 1K$ shows flat, renormalized plateaus which are replaced by uneven 
conductance fluctuations at $T = 50mK$. However a different experiment 
\cite{facer} reveals that a QPC created using similar
split-gate methods shows plateaus which are hardly renormalized at higher 
temperatures, and no conductance fluctuations are seen at lower temperatures. 
This can also be understood from our model: the 
conductance corrections due the junction barriers for a quantum wire are 
gate voltage independent at higher temperatures, but are dependent on it at 
lower temperatures. For the experiment in Ref. \cite{reilly}, $T_l = 0.4K >> 
T=50mK$. Hence, resonance effects are expected at these temperatures. 
This is in contrast to the conductance corrections for a QPC which are
gate voltage independent at all temperatures. In fact, if the quantum wire 
samples of Yacoby {\it et al} have contact regions as long 
as $2 - 6\mu m$ (as found by the authors of Ref. \cite{picciotto} on similar 
samples), this would suggest that their $2\mu m$ long wire is actually 
closer to a quantum point contact. This would help explain the flatness 
of the renormalizations seen over a wide temperature range of $0.3 - 25K$.

We now discuss our attempt to quantitatively understand the variation of 
conductance against temperature as given in the inset of Fig. 3 of the work 
of Yacoby {\it et al} \cite{yacoby}. The conductance given there is 
measured at a fixed value 
of the gate voltage on the plateau of the first sub-band (i.e., close to 
$2g_0$). We find that the conductance correction versus temperature variation 
found by them (i.e., $\delta g \equiv 2g_0 - g$ vs. $T$) is best fitted by a 
function of the form
\beq
\delta g = - 0.3512 ~T^{-0.1058 - 0.0345 T} 
\eeq
as shown in Fig. 3. We find that the goodness of this fit is given by 
the correlation coefficient $R^2 = 0.9955$.
Clearly, this expression for the conductance corrections does not match the 
simple form $\delta g \sim T^{-\alpha}$ given in Sec. $4$ for the QW or QPC 
systems. The presence
of the $T$ dependent piece in the exponent implies that our model is
only qualitatively correct. Several factors could be important in determining 
this complicated temperature dependent power-law. Some of these are:

\begin{itemize}

\item a more extended transition region between the leads and the contacts 
in which the parameters $K$ and $v$ vary smoothly as a function of $x$, 

\item more extended junction barriers lying within the contacts rather than 
the local $\delta$-function barriers that we have studied, and

\item the possibility of the electron-electron interactions having 
a finite range instead of the short-ranged interactions that 
we have used to study our TLL systems.

\end{itemize}

A detailed quantitative comparison of our model with the experiments would, 
therefore, need a more sophisticated treatment taking these factors into 
consideration. We should emphasize here that a temperature and length 
dependence of the conductance correction of the form that we have obtained 
(decreasing at high temperatures or short lengths) 
is a nontrivial effect of the electron interactions, and our simple model has 
already captured this qualitatively. A non-interacting theory does not have 
temperature or length dependences of this kind.

We now discuss the important experimental finding of Liang {\it et al} 
\cite{liang} of the odd-even effect in the transport of electrons through 
a quantum wire in the presence of a magnetic field.
Liang {\it et al} find that as they turn up the external magnetic field (kept 
in plane and aligned along the direction of the channel) from $0$ to $11T$, 
the increasing magnetic field expectedly lifts the spin degeneracy and splits 
each conductance step into two steps, with the heights of both being less than 
$g_0$. Furthermore, at a magnetic field strength of $11T$, they find that the 
difference between the conductance of successive pairs of spin-split sub-bands 
alternates. This shows that the conductance of the odd numbered 
spin-split sub-bands containing the moments aligned with the magnetic field 
undergoes little renormalization (i.e., is close to $g_0$ in their Fig. 4), 
while the conductance of the even numbered spin-split sub-bands containing 
the moments anti-aligned with the magnetic field undergoes a large 
renormalization correction; their Fig. 4 indicates a correction as large 
as $0.3 g_0$. As discussed earlier, this phenomenon can be simply 
understood as the aligned moments seeing a much weaker barrier and the 
anti-aligned moments seeing a much stronger barrier. This is due to the Zeeman 
splitting of the Fermi levels of the up and down spin electrons 
and their interactions with each other. 
Since the difference in renormalizations between the aligned and 
anti-aligned electrons occur for all magnetic fields (i.e., even when the
up and down sub-bands are not spin-split), we suggest the following 
possibility. One can artificially enhance the barrier
strengths so that the difference in renormalizations of the up
and down spins can be made substantial at moderate magnetic fields. More 
importantly, we can vary the gate voltage so as to tune the 
spin polarization with greater transmission to
resonance. This would mean that at these values of the magnetic field
and gate voltage, transmission of one of the polarizations is completely
suppressed and the other one greatly enhanced. This leads us to
the possibility of creating a spin-valve at moderate magnetic fields.

Finally, we comment on a new set of experiments \cite{topinka,crook} which 
have used scanning probe microscopy techniques to study transport through 
QPCs and propose a test for our model based on such a study. In these 
experiments, a negatively charged atomic force microscope tip is held 
at a distance of $100-150nm$ above the 2DEG gas on which the QPC is created 
via split-gate methods. A capacitive coupling between the 2DEG and the 
tip reduces the density of the 2DEG in a small spot directly beneath the tip, 
thereby creating a small depletion region (negatively charged ``bubble") which 
can back-scatter electrons approaching it. The tip then scans the surface 
of the 2DEG reservoir into which the electrons are entering after traveling 
through the QPC, and the two-probe conductance is measured. This allows one 
to ``image" the electron current flowing out from the QPC. Topinka {\it et al} 
\cite{topinka} have made such measurements at a temperature of $T = 1.7K$ and 
find that the electrons flow out into the 2DEG reservoir in streaks from each 
sub-band. The number and nature of the streaks is governed by the 
electron wave function in each sub-band caused by the quantization due to the 
confinement in the transverse direction. They find that the electron flow is 
coherent along 
these streaks quite far from the QPC mouth where they finally disperse into 
the 2DEG. Furthermore, they find that placing the depletion bubble in the
path of a particular streak (at a distance of about $0.3 - 0.5\mu m$ from 
the mouth of the QPC) gives rise to a flat, renormalized plateau only for the 
particular sub-band from which it is emanating, while the other sub-bands give 
the universal conductance value of $2g_0$. This tells us that the effect 
of the gate voltage must vanish quickly since it is not felt beyond distances 
as short as $0.3 - 0.5\mu m$ from the mouth of the QPC. Crook {\it et al} 
\cite{crook} find a series of peaks and troughs upon measuring the 
differential conductance $dg/dV_G$ versus the gate voltage $V_G$ (which are 
caused by the step rises and plateaus for each sub-band respectively) while 
scanning the tip through the QPC. Their finding that the troughs do not fall 
to zero indicate that the conductance corrections caused by the depletion 
bubble (when placed within the QPC) is gate 
voltage dependent as would have been expected. 

Now, the availability of the tip generated depletion bubble as a controlled 
barrier to the flow of electrons through the QPC also 
suggests a possible use of scanning probe microscopy 
techniques to test the predictions of 
our model in a quantitative fashion. This would require the gate voltage to be 
first fixed such that only the lowest sub-band is fully open to the flow 
of electron current, and then the depletion bubble to be placed somewhere on a 
streak emanating from this lowest sub-band at a distance from the QPC mouth; 
the conductance can then be measured by changing the gate 
voltage but holding the temperature fixed. The nature of the 
conductance versus gate voltage curve will tell us whether the gate voltage 
does or does not have any effect on the electrons on the streak at that 
distance from the mouth of the QPC. Furthermore, the gate voltage can then be 
held fixed somewhere on a plateau and the conductance measured as the 
temperature is varied. The form of the conductance corrections versus 
temperature can then be obtained. This entire chain of measurements can then 
be repeated after taking the depletion bubble closer to the QPC mouth and 
into the QPC in a series of steps. Such a series of measurements would 
help answer questions about where the conductance corrections start becoming 
dependent on the gate voltage as well as 
how the conductance corrections vary with temperature when a barrier is 
placed within the QPC or away from the QPC. Such 
experiments could also be carried out with longer QWs to 
check the length dependences of the conductance corrections.

\section{\bf Summary and Outlook}

The main idea in this paper is to introduce a model which explicitly
describes the regions in between the quantum wire and the 2DEG
reservoirs as interacting 1D systems which
are independent of the density of electrons in the quantum wire.
We show that the difference in the strengths of the interactions in
the different regions leads to local junction barriers between the regions; 
the barriers simulate the effects of the imperfect coupling between 
the 2DEG and the quantum wire. Our model leads to the following results for 
wires with no impurities, all of which are in agreement with a large body of 
experimental observations.

\begin{itemize}

\item{} Flat (independent of gate voltage) and uniform (for all the sub-bands) 
renormalizations of the quantized conductance plateaus.

\item{} The renormalizations increase as the temperature is 
lowered or the length of the quantum wire is increased.

\item{} At still lower temperatures, the flatness of the plateaus
disappears and oscillatory features in the conductance can be observed 
which we interpret as resonant transmission through the quantum wire.

\item{} In the presence of a magnetic field, an odd-even effect
is found in the conductance of alternate spin-split sub-bands. This effect 
may be used to construct a spin-valve, which allows only electrons with one 
particular spin to transmit through the wire even if the magnetic field is 
not high enough to completely spin-split the sub-bands.
 
\end{itemize} 

For quantum wires with impurities, which are either intrinsic or
externally imposed as finger gates, the conductance corrections
are always gate voltage dependent and therefore, are neither flat
nor sub-band independent. 

Some interesting questions for future studies include the following.
A quantitative fit to the conductance corrections as a function
of the temperature and wire length still remains to be done. This
would require an even more realistic modeling of the quantum wire
system (including some of the features itemized in the
previous section) as well as more experimental data. 
Theoretical studies at finite frequencies and finite external voltages
across the quantum wire also need to be pursued. 
Finally, one needs to understand several features which are observed
on the rise between two successive plateaus, such as the ``$0.7$ effect''
mentioned in the introduction, the observation of continuous oscillations as 
a function of the gate voltage upon introducing finger gate barriers
\cite{tkachenko}, and the fixed point that exists on the rise as the 
temperature is varied \cite{senz}. For all of these, one needs to study the 
model when some sub-band is partially opened.
 
\vskip .5 true cm
\leftline{\bf Acknowledgments}
\vskip .5 true cm

SL thanks I. Safi for useful correspondence. DS thanks the Council of 
Scientific and Industrial Research, India for financial support through 
grant No. 03(0911)/00/EMR-II.

\vskip .5 true cm

\appendix

\section{\bf Effective action for spinless fermions}

In this Appendix, we will obtain explicitly the $S_0$ part of the effective 
action in terms of the fields $\phi_i,i=1,..4$, at the junctions $x=0,d,l+d$ 
and $x=l+2d=L$ for the $K_L$-$K_C$-$K_W$-$K_C$-$K_L$ model described by the 
action in Eq. (\ref{s0}) by integrating out all degrees of freedom except at 
the positions of the junctions. We will also give the effective action of the
simpler model $K_L$-$K_W$-$K_L$ for comparison, since they have also
not been explicitly given anywhere.

We first start with the simpler $K_L$-$K_W$-$K_C$ model, which is 
defined as a length $L$ quantum wire with interaction parameter $K_W$ 
between $x=0$ and $x=L$, and with leads defined by $K_L=1$ for
$x<0$ and $x>L$, described by the Lagrangian 
\beq
{\cal L}= \int_{-\infty}^0 dx {\cal L}_1 + \int_{0}^{L} 
dx {\cal L}_2 + \int_L^\infty dx {\cal L}_1 ,
\eeq
where
\bea
{\cal L}_1 ~&=&~ {1\over 2K_Lv_L} (\pt\phi)^2 +{v_L\over 2K_L} 
(\px\phi)^2, \nonum \\
{\rm and } \quad 
{\cal L}_2 ~&=&~ {1\over 2K_Wv_W} (\pt\phi)^2 +{v_W\over 2K_W} (\px \phi)^2.
\eea
to set the notation. There 
are three ways to derive the effective action. We can (a) integrate out
the fields at all points in space except at $x=0$ and $L$, or (b) find 
the solution of the equations of motion in terms of the above two fields
and then compute the action from that solution, or (c) compute the 
Green's function $G_\bo (x,x^\prime )$, set $x,x^\prime$ equal
to $0$ or $L$, and invert $G$ to get $S_{eff}$. All the methods 
produce the same result since the original action is purely quadratic.
We will use the second method here because it is technically simpler.

As in other sections, we will work with the 
Euclidean time action for convenience. If all the
fields have a time dependence of the form $\exp (-i\on \tau)$, then 
normalizability of the solutions imply that they should decay exponentially
at $x \rightarrow \pm \infty$. We assume that the solution of the equation
of motion has the following forms in the three regions,
\bea
\tphi(x,\on) &=& \tphi(0,\on)e^{-i\on\tau +|\on|x/v_L}, \quad x<0 \nonum \\
&=& e^{-i\on\tau} (\ttheta_1(\on) e^{|\on| x/v_W} +\ttheta_2(\on)
e^{-\on x/v_W}), \quad 0<x<L \nonum \\
&=& \tphi(L,\on)e^{-i\on\tau+|\on|(L-x)/v_L}, \quad x>L ~.
\eea
Matching solutions at the boundaries $x=0$ and $x=L$ to eliminate $\ttheta_i
(\on)$, and using this solution in the effective action and carrying out the 
spatial integration, we obtain the action 
\bea
S_0 &=& {1\over 2K_L} \sum_{\on} |\on| (\tphi_1^2+\tphi_2^2) 
+ {1\over 2K_W} \sum_{\on} {|\on| \over e^{k_{nW}L} - e^{-k_{nW}L} } 
\times \nonum \\ 
&& \quad \quad \quad [(e^{k_{nW}L}+e^{-k_{nW}L} )(\tphi_1^2+\tphi_2^2) -
4\tphi_1\tphi_2] ,
\eea
where $\tphi_1 \equiv \tphi(0,\on)$ and $\tphi_2 \equiv \tphi(L,\on)$
and $k_{nW}$ and $k_{nL}$ are defined as $|\on|/v_W$ and $|\on|/v_L$ 
respectively. 
In the limit $\on \gg v_L/L, v_W/L$, we get the high frequency effective action 
\beq
S_{high} = {K_L+K_W \over 2K_LK_W} \sum_{\on}|\on|
(\tphi_1^2 +\tphi_2^2) ~,
\eeq
where the two junctions are decoupled as expected. In the low
frequency limit $\on \ll v_L/L, v_W/L$, we get
\beq
S_{low} = {1\over 2K_L} \sum_{\on} |\on| (\tphi_1^2+
\tphi_2^2) + { U_W\over 2} \sum_{\on} |\on| (\tphi_1-\tphi_2)^2 ~,
\eeq
where $U_W = v_W/K_W L$. 

Using the same method as above, we can also obtain the full effective
action for the $K_L$-$K_C$-$K_W$-$K_C$-$K_L$ model in terms of the fields at
the four junctions $\phi_i, i=1,..4$, where $\tphi(0,\on) = \tphi_1(\on) 
\equiv \tphi_1$, $\tphi(d,\on) \equiv \tphi_2$, $\tphi(l+d,\on)
\equiv \tphi_3$ and $\tphi(L=l+2d,\on) \equiv
\tphi_4$. The solutions in the five regions can be written as
\bea
\tphi(x, \on) &=& \tphi_1 e^{k_{nL}x}, \quad x<0 \nonum \\
&=& B e^{k_{nC}x} + Ce^{-k_{nC}x}, \quad 0<x<d \nonum \\
&=& D e^{k_{nW}x} + Ee^{-k_{nW}x}, \quad d<x<l+d \nonum \\
&=& F e^{k_{nC}x} + Ge^{-k_{nC}x}, \quad l+d<x<L \nonum \\
&=& \tphi_4 e^{k_{nL}(L-x)}, \quad x<0 ~.
\eea
where by matching the solutions at $x=0,d,l+d$ and $L$, we can
obtain the functions of $\on$, $B,C,D,E,F$ and $G$ in terms of the
$\tphi_i, i=1...4$, and $k_{nC}$ is defined as $k_{nC} = |\on| /v_C$. 
Substituting this solution in the 
action and integrating over all space, we get the effective action
\bea
S_{eff} &=& \sum_{\on}\{{1\over 2K_L} |\on| (\tphi_1^2+\tphi_4^2) \nonum \\
&& +{1\over 2K_C} |\on| [(B^2+G^2e^{-2k_n L}) (e^{2k_n d} - 1) + (C^2+F^2
e^{2k_nL})(1-e^{-2k_n d})] \nonum \\
&& +{1\over 2K_W} [|\on| [D^2(e^{2k_n (L-d)} - e^{2k_n d}) -
E^2(e^{-2k_n (L-d)} - e^{-2k_n d})] \}~,
\eea
with
\bea
B={\tphi_2 -\tphi_1e^{-k_{nC}d}\over e^{k_{nC} d} -e^{-k_{nC}d}}, \quad && C=
{\tphi_1e^{k_{nC}d}-\tphi_2\over e^{k_{nC} d} -e^{-k_{nC}d}}\nonum \\
D={\tphi_3 -\tphi_2e^{-k_{nW}l}\over e^{k_{nC}d} (e^{k_{nW}l} -e^{-k_{nW}l})}, 
\quad && E= {\tphi_2e^{k_{nW}l}-\tphi_2\over e^{-k_{nC}d}
(e^{k_{nW} l} -e^{-k_{nW}l})}\nonum \\
F={\tphi_4 -\tphi_3e^{-k_{nC}d}\over e^{k_{nC}d +k_{nW}l}
(e^{k_{nC} d} -e^{-k_{nC}d})}, 
\quad && G= {\tphi_3e^{k_{nC}d}-\tphi_4\over e^{-k_{nC}d -k_{nW}l}
(e^{k_{nC} d} -e^{-k_{nC}d})} ~. 
\eea
The action finally simplifies to 
\bea
S_{eff} &=& {1\over 2K_L} \sum_{\on} |\on| (\tphi_1^2+\tphi_4^2) \nonum \\
&&\hspace*{-0.2cm} +{1\over 2K_C} \sum_{\on} {|\on| \over e^{k_{nC}d} 
- e^{-k_{nC}d}} [(e^{k_{nC}d} +e^{-k_{nC}d}) (\tphi_1^2 +
\tphi_2^2 + \tphi_3^2+\tphi_4^2) -4\tphi_1\tphi_2 -4\tphi_3\tphi_4] \nonum \\
&&\hspace*{-0.2cm} +{1\over 2K_W} \sum_{\on} {|\on| \over e^{k_{nW}l} 
- e^{-k_{nW}l}} [(e^{k_{nW}l} +e^{-k_{nW}l}) (\tphi_2^2+\tphi_3^2) 
-4\tphi_2\tphi_3] ~,
\eea
where $\tphi_1 \equiv \tphi(0,\on)$ and $\tphi_2 \equiv \tphi(L,\on)$.
The high and low frequency limits of this effective
action have been used in Sec. 3, to compute the finite temperature and
finite length corrections off-resonance.

\section{\bf Effective action for spinful fermions}

In this section, we will explicitly compute the effective action for 
spinful fermions in the $K_L$-$K_C$-$K_W$-$K_C$-$K_L$ model. Although,
the method followed is exactly the same as that in the previous
section for spinless fermions, we do it explicitly because there are 
a few points where the inclusion of spin makes a difference.

The effective action for spinful fermions is normally computed in
terms of the `charge' and `spin' field variables defined as $\phi_\rho= 
(\phi_\ua +\phi_\da)/\sqrt{2}$ and $\phi_\sigma =(\phi_\ua
-\phi_\da)/\sqrt{2}$ because in the presence of interactions, the spin $\ua$
and $\da$ fermions are mixed (remember the Hubbard term ${\cal U} \sum_i
n_{i\ua} n_{i\da}$). Here, in our model with contacts, the interaction
term $\cal U$ is different in the contact region and in the wire
region. But since the linear combination that diagonalizes the interaction 
is independent of the value of $\cal U$, the action in terms of the 
$\phi_\rho$ and $\phi_\sigma$ fields are decoupled. In the 
presence of a magnetic
field, in the next Appendix, we will see that the action continues to be
diagonalizable; however, the diagonal fields are defined in terms of
mixing angles which explicitly depend on $\cal U$ and the magnetic
field and hence are different in the leads, the contacts and the wire. 

The starting action for the spinful fermions is given in
Eqs. (\ref{s0}) and (\ref{sol}) in the text in terms of the charge and
spin fields. As in the earlier Appendix, we will obtain the solution of the 
equations of motion in terms of the eight fields $\tphi_{ia}, i=1...4,
a=\rho , \sigma$ defined to be at the positions $x=0,d,l+d$ and $L=l+2d$ and
then compute the effective action from that solution. We 
assume that the solutions in the five regions can be written as
\bea
\tphi_{a}(x, \on) &=& \tphi_{1a} e^{k_{nLa}x}, \quad x<0 \nonum \\
&=& B_a e^{k_{nCa}x} + C_ae^{-k_{nCa}x}, \quad 0<x<d \nonum \\
&=& D_a e^{k_{nWa}x} + E_ae^{-k_{nWa}x}, \quad d<x<l+d \nonum \\
&=& F_a e^{k_{nCa}x} + G_ae^{-k_{nCa}x}, \quad l+d<x<L \nonum \\
&=& \tphi_{4a} e^{k_{nL}(L-x)}, \quad x<0 ~.
\eea
and as before, the coefficients, $B_a,C_a,D_a,E_a,F_a$ and $G_a$ can
be found in terms of the $\tphi_{ia}, i=1...4, a=\rho , \sigma$
by matching the solutions at $x=0,d,l+d$ and $L$. Note that 
$k_{nWa}$, $k_{nCa}$ and $k_{nL}$ are defined as $|\on|/v_{Wa}$, 
$|\on|/v_{Ca}$ and $|\on|/v_{La}$ respectively. Substituting this solution 
in the action and integrating over all space, we get the effective action as
\bea
S &=& \sum_{a=\rho , \sigma}\{{1\over 2K_{La}} 
\sum_{\on} |\on| (\tphi_{1a}^2+\tphi_{4a}^2) + {1\over 2K_{Ca}} 
\sum_{\on} {|\on| \over e^{k_{nCa}d} - e^{-k_{nCa}d}} \times\nonum \\ 
&& \quad \quad \quad [(e^{k_{nCa}d} +e^{-k_{nCa}d}) (\tphi_{1a}^2+
\tphi_{2a}^2 + \tphi_{3a}^2+\tphi_{4a}^2) -4\tphi_{1a}\tphi_{2a}
-4\tphi_{3a}\tphi_{4a}] \nonum \\
&+& {1\over 2K_{Wa}} \sum_{\on} {|\on| \over e^{k_{nWa}l} - e^{-k_{nWa}l}} 
[(e^{k_{nWa}l} +e^{-k_{nWa}l}) (\tphi_{2a}^2+\tphi_{3a}^2)
-4\tphi_{2a} \tphi_{3a}]\}.
\label{spineff} 
\eea
The high and low frequency limits of this effective
action have been used in Sec. 3, to discuss the various resonances that
are possible in the low temperature limit and to explicitly 
compute the off-resonance corrections to the conductances at 
finite temperatures and for finite length wires.

\section{\bf Effective action for spinful electrons in the
presence of a magnetic field}

We present here the calculation for the effective action for
spinful fermions in a magnetic field 
when the mixing angle $\gamma$ is the same in both the contacts as well as 
the QW, i.e., the short ranged electron-electron interaction $\cal U$ is equal 
in all the three TLLs. We start with the action given in Eq. (\ref{actmag}) 
in Sec. 5 and will now integrate out the fields at all points except at the
four junctions as these will be 
the sites for the two outer barriers while the two inner junctions are the 
ends of the region to which the gate voltage couples. Thus, we write down 
the equations of motion in each of the five regions and solve them. If all the
fields have a time dependence of the form $\exp (-i\on \tau)$, then 
normalizability of the solutions imply that they should decay exponentially
at $x \rightarrow \pm \infty$. The general solution is given by
\bea
\tphi_\ua(x) &=& A_\ua e^{k_\ua x}, \quad \tphi_\da(x) = A_\da e^{k_\da x}, 
\quad x<0 \nonum \\
\tphi_\pm(x) &=& B_{\pm}e^{k_\pm x}+C_{\pm}e^{-k_\pm x}, \quad 0<x<d \nonum \\
\tphi_\pm(x) &=& D_{\pm}e^{\tk_\pm x}+E_{\pm}e^{-\tk_\pm x}, \quad d<x<l+d 
\nonum \\ 
\tphi_\pm(x) &=& F_{\pm}e^{k_\pm x}+G_{\pm}e^{-k_\pm x}, \quad l+d<x<l+2d 
\nonum \\
\tphi_\ua(x) &=& H_\ua e^{k_\ua x}, \quad \tphi_\da = H_\da e^{k_\da x}, \quad
x<0 ~,
\label{coeff}
\eea
where we have defined $\tk_\pm = |\on| /v_{W\pm}$, $k_\pm = |\on| /
v_{C\pm}$, $k_\ua = |\on|/v_{F\ua}$ and $k_\da = |\on|/v_{F\da}$.

We now solve for the coefficients $A,...,H$ in Eq. (\ref{coeff})
by matching the fields
$\tphi_\ua$ and the $\tphi_\da$ at the four junctions. At this point, we make
the simplifying assumption that the mixing angles $\gamma_C$ and $\gamma_W$
(defined as in Eqs. (\ref{transf1}) and (\ref{transf2})) in the contact and 
wire regions are equal to each other, $\gamma_C = \gamma_W = \gamma$. 
This implies that $\tphi_{W\pm} = \sqrt{{K_{W\pm} v_{W\pm}} \over {K_{C\pm} 
v_{C\pm}}} \tphi_{C\pm}$ at $x=d$ and $x=l+d$.
We find that the coefficients are given by 
\bea
A_\ua &=& \tphi_{1\ua}, \quad A_\da = \tphi_{1\da} \nonum \\
B_{\pm} &=& {\tphi_{2\pm} - \tphi_{1\pm}
e^{-k_\pm d}\over D_{C\pm}} \nonum \\
C_{\pm} &=&- {\tphi_{2\pm} + \tphi_{1\pm}
e^{k_\pm d}\over D_{C\pm}} \nonum \\
D_{\pm} &=& \sqrt{{K_{W\pm} v_{W\pm}} \over {K_{C\pm} v_{C\pm}}}
e^{-\tk_\pm d} ~{(-\tphi_{2\pm} 
e^{-\tk_\pm l} + \tphi_{3\pm}) \over D_{W\pm}} 
\nonum \\
E_{\pm} &=& \sqrt{{K_{W\pm} v_{W\pm}} \over {K_{C\pm} v_{C\pm}}} 
e^{\tk_\pm d}~{(\tphi_{2\pm} 
e^{\tk_\pm l} - \tphi_{3\pm}) \over D_{W\pm}} 
\nonum \\
F_{\pm} &=& -{\tphi_{3\pm}e^{-k_\pm(l+2d)} - \tphi_{4\pm} 
e^{-k_\pm d}\over D_{C\pm}} \nonum \\
G_{\pm} &=& {\tphi_{3\pm}e^{k_\pm(l+2d)} - \tphi_{4\pm}
e^{k_\pm d}\over D_{C\pm}} \nonum \\
H_{\ua} &=& \tphi_{4\ua}e^{k_\ua(l+2d)}, \quad H_\da = \tphi_{4\da}
e^{k_\da(l+2d)} ~,
\eea
where
\bea
D_{C\pm} &=& e^{k_\pm d} - e^{-k_\pm d} ~ \nonum \\
{\rm and}\quad D_{W\pm} &=& e^{\tk_\pm l} - e^{-\tk_\pm l} ~.
\eea

Then using the relations written down in Eq. (\ref{transf1})
connecting the $\phi_{\pm}$ and $\phi_{\ua,\da}$ fields, 
we find that the effective Lagrangian density is given by
\bea
{\cal L} = && \frac{|\on|}{2K_L}(\tphi_{1\ua}^2 + \tphi_{1\da}^2 
\tphi_{4\ua}^2 + \tphi_{4\da}^2) \nonum \\
&& + \frac{|\on|}{2}[(p^2\frac{N_{C+}} {K_{C+}D_{C+}} + r^2\frac{N_{C-}}
{K_{C-}D_{C-}})(\tphi_{1\ua}^2 + \tphi_{2\ua}^2 
+ \tphi_{3\ua}^2 + \tphi_{4\ua}^2) \nonum \\
&& + (q^2\frac{N_{C+}} {K_{C+}D_{C+}} + s^2\frac{N_{C-}}
{K_{C-}D_{C-}})(\tphi_{1\da}^2 
+ \tphi_{2\da}^2 + \tphi_{3\da}^2 + \tphi_{4\da}^2) \nonum \\
&& - 4(\frac{p^2}{K_{C+}D_{C+}} + 
\frac{r^2}{K_{C-}D_{C-}})(\tphi_{1\ua}\tphi_{2\ua} 
+ \tphi_{3\ua}\tphi_{4\ua}) \nonum \\
&& - 4(\frac{q^2}{K_{C+}D_{C+}} + \frac{s^2}{K_{C-}D_{C-}})
(\tphi_{1\da} \tphi_{2\da} + \tphi_{3\da}\tphi_{4\da})] \nonum \\
&& + |\on| [(\frac{pq N_{C+}}{K_{C+} D_{C+}} 
+ \frac{rs N_{C-}}{K_{C-} D_{C-}})
(\tphi_{1\ua}\tphi_{1\da} + \tphi_{2\ua}\tphi_{2\da} 
+ \tphi_{3\ua}\tphi_{3\da} + \tphi_{4\ua}\tphi_{4\da}) \nonum \\
&& - 2(\frac{pq}{K_{C+}D_{C+}} + \frac{rs}
{K_{C-}D_{C-}})(\tphi_{1\ua}\tphi_{2\da} + 
\tphi_{1\da}\tphi_{2\ua} + \tphi_{3\ua}\tphi_{4\da} 
+ \tphi_{3\da}\tphi_{4\ua})] \nonum \\
&& + \frac{|\on|}{2}[(p^2\frac{N_{W+}} {K_{W+}D_{W+}} + r^2\frac{N_{W-}}
{K_{W-}D_{W-}})(\tphi_{2\ua}^2 + \tphi_{3\ua}^2) \nonum \\
&& + (q^2\frac{N_{W+}} {K_{W+}D_{W+}} + s^2\frac{N_{W-}}
{K_{W-}D_{W-}})(\tphi_{2\da}^2 + \tphi_{3\da}^2) \nonum \\
&& - 4(\frac{p^2}{K_{W+}D_{W+}} +
\frac{r^2}{K_{W-}D_{W-}})\tphi_{2\ua}\tphi_{3\ua}
\nonum \\
&& - 4(\frac{q^2}{K_{W+}D_{W+}} + 
\frac{s^2}{K_{W-}D_{W-}})\tphi_{2\da}\tphi_{3\da}] \nonum \\
&& + |\on| [(\frac{pq N_{W+}}{K_{W+} D_{W+}} + \frac{rs N_{W-}}{K_{W-}D_{W-}})
(\tphi_{2\ua}\tphi_{2\da} + \tphi_{3\ua}\tphi_{3\da}) \nonum \\
&& - 2(\frac{pq}{K_{W+}D_{W+}} + \frac{rs}
{K_{W-}D_{W-}})(\tphi_{2\ua}\tphi_{3\da} + \tphi_{2\da}\tphi_{3\ua})],
\eea
where
\bea
N_{C\pm} &=& e^{k_\pm d} + e^{-k_\pm d} ~, \nonum \\
N_{W\pm} &=& e^{\tk_\pm l} + e^{-\tk_\pm l} ~,
\eea
and $D_{C\pm}, D_{W\pm}, k_{\pm}$ and $\tk_{\pm}$ have already been 
defined above.

\section{\bf Calculation of the Green's function in our model for the 
Quantum Wire}

Here, we will present a calculation of the Green's function for the bosonic 
excitations in the model that we have presented for the quantum wire system 
of spinless fermions. The method we follow is along the lines of the 
calculation presented by Maslov and Stone \cite{maslov}. We will study 
the case when there are no barriers present anywhere in the system. Then, 
we see that the Euclidean action $S_E$ in all the five distinct TLL regions 
in our model (Fermi lead, contact, QW, contact and Fermi lead) is given by
\beq
S_E = \frac{1}{2} \int_{0}^{\beta} d\tau \int_{-\infty}^{\infty}
[\frac{1}{K(x)v(x)}(\partial_{\tau}\phi)^2 + 
\frac{v(x)}{K(x)}(\partial_{x}\phi)^2],
\eeq
with $K(x) = K_L, v(x) = v_L$ in the first and fifth (Fermi lead) regions, 
$K(x) = K_C, v(x) = v_C$ in the second and fourth (contact) regions and 
$K(x) = K_W, v(x) = v_W$ in the third (QW) region. Then, defining the 
two-point bosonic Green's function/propagator (in Euclidean time $\tau$) as
\beq
G(x,x',\tau) = <T_{\tau}\phi(x,\tau)\phi(x',0)>,
\eeq
it can be shown that the equation satisfied by the Fourier transform of 
the above Green's function $G_{\bar{\omega}}(x,x')$ is
\beq
\{-\partial_{x}(\frac{v(x)}{K(x)}\partial_{x}) + 
\frac{1}{K(x)v(x)}\bar{\omega}^2\} G_{\bar{\omega}}(x,x') = \delta(x-x').
\eeq
We now have to solve the above equation to obtain a functional form for 
$G_{\bar{\omega}}(x,x')$. We know that the interaction parameter $K$ and 
the velocity $v$ change abruptly at each of the junctions and that the 
two Fermi leads are semi-infinite in length (i.e. $G_{\bar{\omega}}(x,x')$ 
must decay to zero as $x\rightarrow \pm\infty$). As we are interested in 
finding the one-point Green's function at a point in the left contact, we 
will choose $x'$ to lie between $0$ (the left lead-contact junction) and 
$d$ (the left contact-QW junction). Furthermore, we know that the 
Green's function $G_{\bar{\omega}}(x,x')$ must satisfy the following 
boundary conditions: (a) $G_{\bar{\omega}}(x,x')$ must be continuous at 
$x = 0, x', d, l+d$ and $l+2d$ (b) $(\frac{v(x)}{K(x)})\partial_{x}
G_{\bar{\omega}}(x,x')$ must be continuous at $x = 0, d, l+d$ and $l+2d$ 
and 
\beq
- \frac{v(x)}{K(x)}\partial_{x}G_{\bar{\omega}}(x,x')
\vert_{x=x'-0}^{x=x'+0} = 1,
\eeq
i.e., $(\frac{v(x)}{K(x)})\partial_{x}G_{\bar{\omega}}(x,x')$ undergoes a 
jump of unity at $x=x'$. It is then easily seen that the solution for 
$G_{\bar{\omega}}(x,x')$ is of the form
\bea
G_{\bar{\omega}}(x,x') &=& A e^{\vert\bar{\omega}\vert x/v_L} 
\quad \quad \quad \quad ~~~~~~~~~\textrm{for $x\leq 0$} \nonum \\
&=& B e^{\vert\bar{\omega}\vert x/v_C} + 
C e^{-\vert\bar{\omega}\vert x/v_C} \quad ~\textrm{for $0<x\leq x'$} \nonum \\
&=& D e^{\vert\bar{\omega}\vert x/v_C} + 
E e^{-\vert\bar{\omega}\vert x/v_C} \quad ~\textrm{for $x'<x\leq d$} \nonum \\
&=& F e^{\vert\bar{\omega}\vert x/v_C} + 
G e^{-\vert\bar{\omega}\vert x/v_C} \quad ~\textrm{for $d<x\leq l+d$} \nonum \\
&=& H e^{\vert\bar{\omega}\vert x/v_C} + 
I e^{-\vert\bar{\omega}\vert x/v_C} \quad ~~\textrm{for $l+d<x\leq l+2d$} 
\nonum \\
&=& J e^{-\vert\bar{\omega}\vert x/v_C} \quad \quad \quad \quad 
~~~~~~~~\textrm{for $l+2d<x$} ~.
\eea 
The coefficients $A, B, \ldots , J$ are found by matching the boundary 
conditions. To begin with, it is worth noting that in the dc limit 
of $\bar{\omega}\rightarrow 0$, we find that 
\beq
A = B + C = D + E = F + G = H + I = J = \frac{K_L}{2\vert\bar{\omega}\vert},
\eeq
which gives the dc conductance to be
\beq
g = \frac{2e^2}{h} K_L = 2g_0 K_L.
\eeq
This gives the perfect quantized conductance observed in several experiments 
on transport of electrons through a QPC when we take 
the leads to be Fermi liquids with $K_L = 1$. 

We now give the expressions for the Green's functions for the case when 
both $x$ and $x'$ are taken equal to $a$ at a point in the left contact:
\bea
&&\hspace{-1cm}G_{\bar{\omega}}(a,a)= \nonum \\
&&\hspace{-1cm}\frac{K_C}{2\vert\bar{\omega}\vert}
\frac{\{(r-p-q-s)e^{\vert\bar{\omega}\vert d/v_C}(1 + 
\gamma_1 e^{-\vert\bar{\omega}\vert 2a/v_C}) - 
(r+p+q-s)e^{-\vert\bar{\omega}\vert d/v_C}
(e^{\vert\bar{\omega}\vert 2a/v_C} + \gamma_1)\}}
{\{(r-p-q-s)e^{\vert\bar{\omega}\vert d/v_C} + 
\gamma_1 (r+p+q-s)e^{-\vert\bar{\omega}\vert d/v_C}\}} , \nonum \\
&&
\eea
where
\bea
p &=& e^{-\vert\bar{\omega}\vert l(\frac{1}{v_C}+\frac{1}{v_W})}
(\gamma_1 e^{-2\vert\bar{\omega}\vert d/v_C} (1+\frac{K_W}{K_C}) 
+ (1-\frac{K_W}{K_C})) \nonum \\
q &=& e^{-\vert\bar{\omega}\vert l(\frac{1}{v_C}-\frac{1}{v_W})}
(\gamma_1 e^{-2\vert\bar{\omega}\vert d/v_C} (1-\frac{K_W}{K_C}) 
+ (1+\frac{K_W}{K_C})) \nonum \\
r &=& e^{-\vert\bar{\omega}\vert l(\frac{1}{v_C}+\frac{1}{v_W})}
(\gamma_1 e^{-2\vert\bar{\omega}\vert d/v_C} (1+\frac{K_C}{K_W}) 
+ (\frac{K_C}{K_W}-1)) \nonum \\
s &=& e^{-\vert\bar{\omega}\vert l(\frac{1}{v_C}-\frac{1}{v_W})}
(\gamma_1 e^{-2\vert\bar{\omega}\vert d/v_C} (\frac{K_C}{K_W}-1) 
+ (1+\frac{K_C}{K_W})) \nonum \\
\gamma_1 &=& \frac{K_L - K_C}{K_L + K_C}.
\eea
The results upon taking the limits corresponding to the various 
frequency (or temperature) regimes are given in the section 
where the conductance is computed for quantum wires and quantum 
point contacts with a junction barrier in the left contact region and we
will not repeat them here. 

We also give the general form of the two-point propagator 
$G_{\bar{\omega}}(x,y)$ for when $x$ is a point in the right lead 
and $y$ is a point in the left contact: 
\beq
G_{\bar{\omega}}(x,y) = \frac{2K_2}{\vert\bar{\omega}\vert} 
\frac{e^{\vert\bar{\omega}\vert (l+2d)(\frac{1}{v_L}-\frac{1}{v_C})} 
e^{\vert\bar{\omega}\vert \frac{d}{v_C}} 
(1+\gamma_1)^{2} e^{\vert\bar{\omega}\vert (\frac{y}{v_C}- \frac{x}{v_L})}}
{\{(p+q+s-r)e^{\vert\bar{\omega}\vert d/v_C} +
\gamma_1 (s-r-p-q)e^{-\vert\bar{\omega}\vert d/v_C}\}} ~,
\eeq
where the expressions for $p, q, r, s$ and $\gamma_1$ have already been 
given earlier.

Now, we give the expression for the one-point propagator at a 
point $a$ inside the quantum wire:
\beq
G_{\bar{\omega}}(a,a)= \frac{K_W}{2\vert\bar{\omega}\vert}
\frac{(je^{\vert\bar{\omega}\vert a/v_W} + 
ke^{-\vert\bar{\omega}\vert a/v_W}) (me^{\vert\bar{\omega}\vert a/v_W} +
ne^{-\vert\bar{\omega}\vert a/v_W})} {(jn - km )} ,
\eeq
where 
\bea
j &=& (1+\frac{K_W}{K_C})e^{\vert\bar{\omega}\vert 
d(\frac{1}{v_C}-\frac{1}{v_W})} + 
\gamma_1 (1-\frac{K_W}{K_C})e^{-\vert\bar{\omega}\vert 
d(\frac{1}{v_C}+\frac{1}{v_W})} \nonum \\
k &=& (1-\frac{K_W}{K_C})e^{\vert\bar{\omega}\vert 
d(\frac{1}{v_C}+\frac{1}{v_W})} +
\gamma_1 (1+\frac{K_W}{K_C})e^{-\vert\bar{\omega}\vert
d(\frac{1}{v_C}-\frac{1}{v_W})} \nonum \\
m &=& (1-\frac{K_W}{K_C})e^{-\vert\bar{\omega}\vert
(l+d)(\frac{1}{v_C}+\frac{1}{v_W})} +
\gamma_1 (1+\frac{K_W}{K_C})e^{-\vert\bar{\omega}\vert
(l+d)(\frac{1}{v_C}+\frac{1}{v_W})}e^{-2\vert\bar{\omega}\vert
\frac{d}{v_C}} \nonum \\
n &=& (1+\frac{K_W}{K_C})e^{-\vert\bar{\omega}\vert
(l+d)(\frac{1}{v_C}-\frac{1}{v_W})} + 
\gamma_1 (1-\frac{K_W}{K_C})e^{-\vert\bar{\omega}\vert
(l+d)(\frac{1}{v_C}-\frac{1}{v_W})}e^{-2\vert\bar{\omega}\vert \frac{d}{v_C}}.
\eea 
Again, we will not give the results of taking the various limits 
corresponding to the different frequency (or temperature) regimes 
as these have already been quoted in the section on the conductance 
of a quantum wire and quantum point contact.

\newpage

\noindent {\bf Figure Captions}
\vskip 1 true cm

\noindent {1.} Schematic diagram of the model showing the lead regions 
(marked FL for Fermi liquid), the contact regions (C) of length $d$, and the 
quantum wire (QW) of length $l$. The interaction parameters in these three 
regions are denoted by $K_L$, $K_C$ and $K_W$ respectively.

\noindent {2.} Diagram showing the Fermi energy $E_F$ (with a thermal spread 
of $k_B T$) in relation to the sub-bands within the quantum wire. The 
conductance will lie on a plateau if the energy levels are as shown in (a),
while the conductance will be at a step between one plateau and the next if 
the energy levels are as shown in (b).

\noindent {3.} A plot showing the curve fitted to the conductance corrections 
$\delta g$ versus temperature $T$ data obtained from the inset of Fig. 3 of 
Yacoby {\it et al} \cite{yacoby}. The expression for the curve and the value 
of the correlation coefficient are given in the text.

\newpage

\begin{figure}
\begin{center}
\epsfig{figure=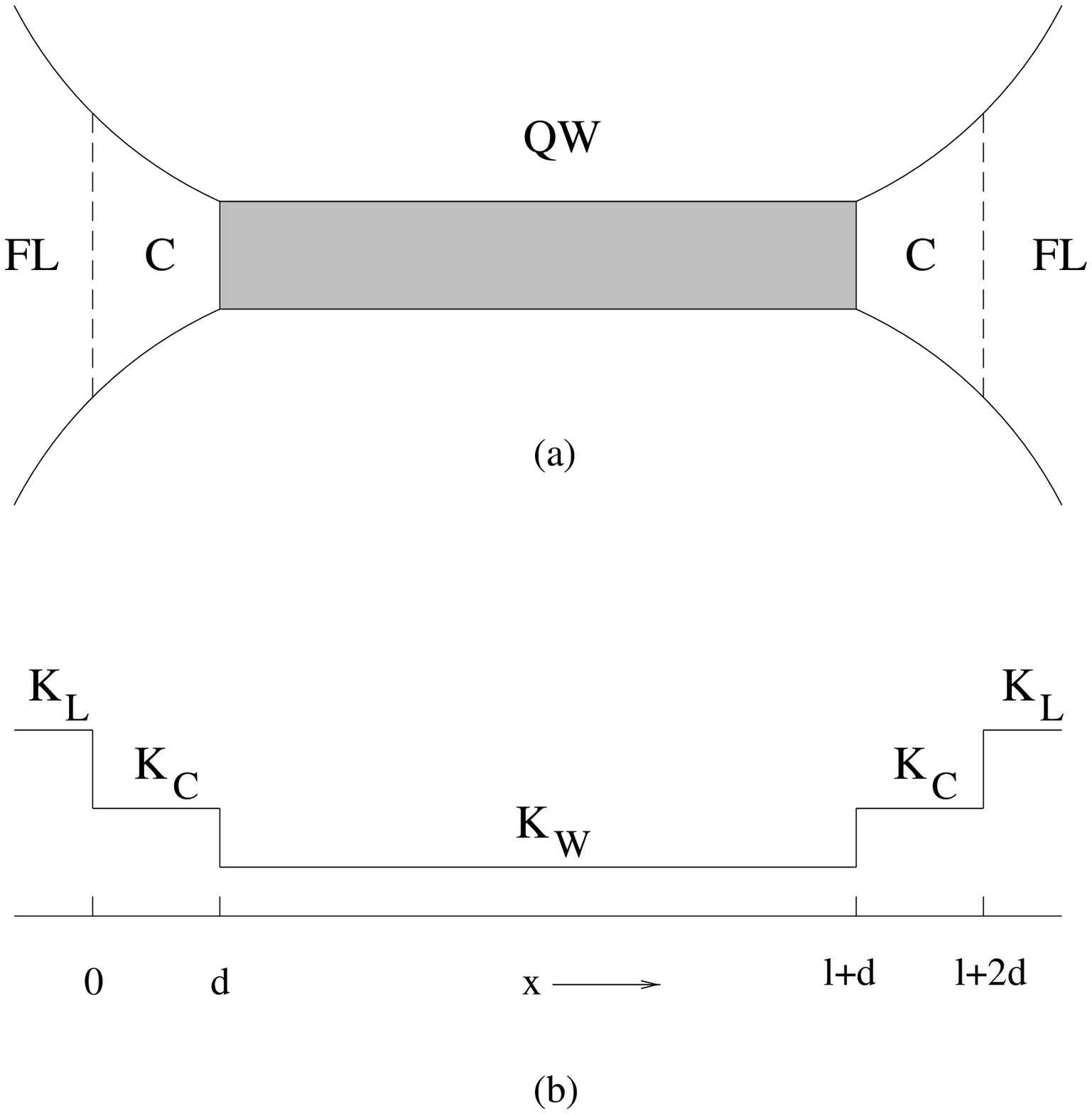,bbllx=50,bblly=0,bburx=550,bbury=800,height=21cm}
\end{center}
\vspace*{-3 cm}
\centerline{Fig. 1}
\end{figure}

\begin{figure}
\begin{center}
\epsfig{figure=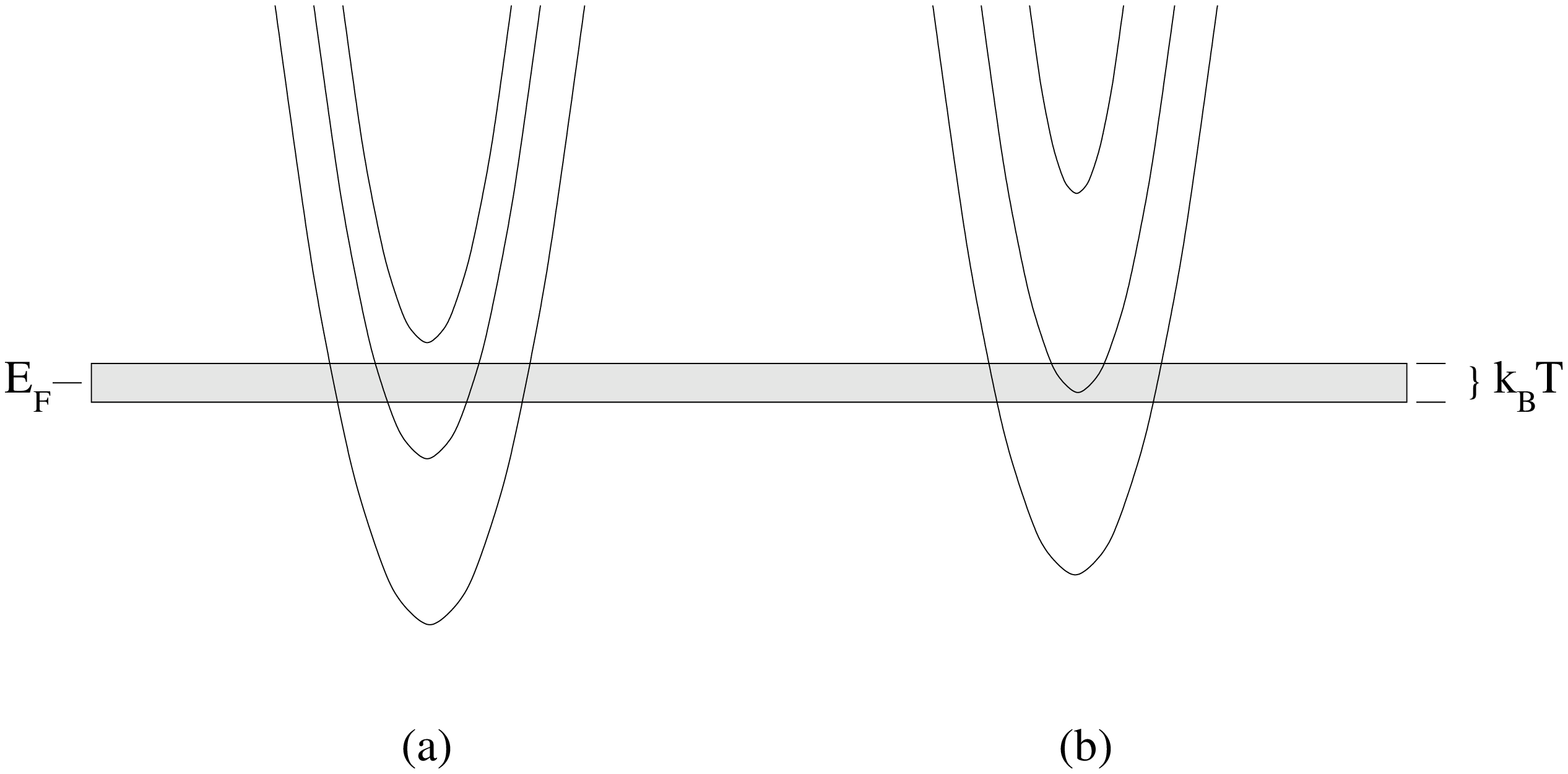,angle=-90,width=16cm}
\end{center}
\vspace*{2 cm}
\centerline{Fig. 2}
\end{figure}

\begin{figure}
\begin{center}
\epsfig{figure=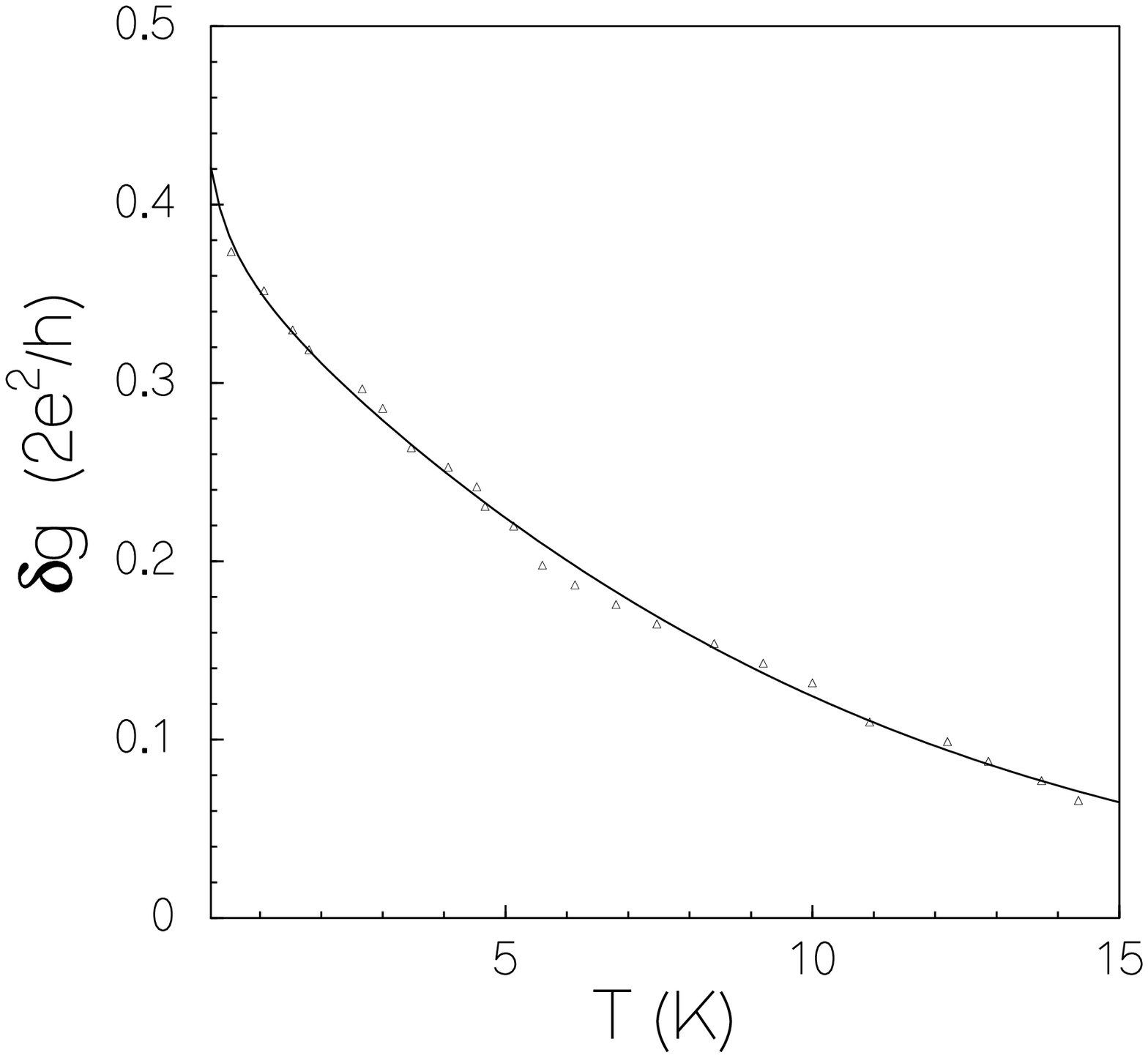,bbllx=50,bblly=-200,bburx=550,bbury=600,height=21cm}
\end{center}
\vspace*{-5 cm}
\centerline{Fig. 3}
\end{figure}

\end{document}